\documentclass[12pt]{JHEP3}
\def\bra#1{\left\langle #1\right|}
\def\ket#1{\left| #1\right\rangle}
\def\beq{\begin{equation}}
\def\eeq{\end{equation}}
\def\bea{\begin{eqnarray}}
\def\eea{\end{eqnarray}}
\def\nn{\nonumber}

\def\sss{\scriptscriptstyle}
\def\Bbar{{\overline B}^0}
\def\bd{B_d^0}

\def\bs{B_s^0}
\def\bsbar{{\overline{B_s^0}}}
\def\ks{K_{\sss S}}
\def\kbar{{\bar K}^0}
\def\Auc{{\cal A}_{uc}}
\def\Atc{{\cal A}_{tc}}
\def\Aut{{\cal A}_{ut}}
\def\Act{{\cal A}_{ct}}
\def\Autp{{\cal A}'_{ut}}
\def\Actp{{\cal A}'_{ct}}
\def\Atcp{{\cal A}'_{tc}}

\def\aI{a_{\sss I}}
\def\aR{a_{\sss R}}

\def\tildeM{{\widetilde M}}
\def\tildeAut{{\widetilde{\cal A}}_{ut}}
\def\tildeAct{{\widetilde{\cal A}}_{ct}}
\def\tildeAtc{{\widetilde{\cal A}}_{tc}}
\def\tildeaI{{\widetilde a}_{\sss I}}
\def\tildeaR{{\widetilde a}_{\sss R}}
\def\tildeB{{\widetilde B}}
\def\tildeActp{{\widetilde{\cal A}}'_{ct}}
\def\tildeAtcp{{\widetilde{\cal A}}'_{tc}}
\def\ssbar{(s{\bar s})}
\def\pew{P_{\sss EW}}
\def\pewc{P_{\sss EW}^{\sss C}}
\def\pewcp{P_{\sss EW}^{\prime \sss C}}
\def\tildeT{{\widetilde T}}
\def\tildeE{{\widetilde E}}
\def\tildeP{{\widetilde P}}
\def\tildeA{{\widetilde A}}
\def\tildePA{{\widetilde{PA}}}

\def\tildepewc{{\widetilde P}_{\sss EW}^{\sss C}}
\def\tildepewcp{{\widetilde P}_{\sss EW}^{\prime \sss C}}
\def\btod{{\bar b} \to {\bar d}}
\def\btos{{\bar b} \to {\bar s}}

\def\Vud{V_{ub}^*V_{ud}}
\def\Vus{V_{ub}^*V_{us}}
\def\Vcd{V_{cb}^*V_{cd}}
\def\Vcs{V_{cb}^*V_{cs}}

\def\epjc#1#2#3{{ Eur.\ Phys.\ J.}\ {\bf C#1}, #3 (#2)}
\def\jhep#1#2#3{{\it JHEP}\ {\bf #1}, #3 (#2)}

\def\npb#1#2#3{{ Nucl.\ Phys.} {\bf B#1}, #3 (#2)}
\def\plb#1#2#3{{ Phys.\ Lett.} {\bf #1B}, #3 (#2)}
\def\prd#1#2#3{{ Phys.\ Rev.} {\bf D#1}, #3 (#2)}
\def\newprd#1#2#3{{ Phys.\ Rev.} {\bf D#1}, #3 (#2)}
\def\prl#1#2#3{{ Phys.\ Rev.\ Lett.} {\bf #1}, #3 (#2)}

\def\zpc#1#2#3{{ Zeit.\ Phys.} {\bf C#1}, #3 (#2)}

\title{Obtaining CKM Phase Information from {\boldmath $B$} Penguin Decays}

\author{Alakabha Datta\\ 
Department of Physics, University of Toronto, \\
\qquad 60 St.\ George Street, Toronto, ON, Canada M5S 1A7 \\
E-mail: \email{datta@physics.utoronto.ca}}

\author{David London\\
Physics Department, McGill University,\\
\qquad 3600 University St., Montr\'eal QC, Canada H3A 2T8; \\
Laboratoire Ren\'e J.-A. L\'evesque, Universit\'e de Montr\'eal,\\
\qquad C.P. 6128, succ. centre-ville, Montr\'eal, QC, Canada H3C 3J7 \\
E-mail: \email{london@lps.umontreal.ca}}

\abstract{We discuss a method for extracting CP phases from pairs of
$B$ decays which are related by flavor SU(3). One decay ($B^0 \to M_1
M_2$) receives a significant $\btod$ penguin contribution. The second
($B' \to M'_1 M'_2$) has a significant $\btos$ penguin contribution,
but is dominated by a single amplitude. CP phase information is
obtained using the fact that the $B' \to M'_1 M'_2$ amplitude is
related by SU(3) to a piece of the $B^0 \to M_1 M_2$ amplitude. The
leading-order SU(3)-breaking effect ($\sim 25\%$) responsible for the
main theoretical error can be removed. For some decay pairs, it can be
written in terms of known decay constants. In other cases, it involves
a ratio of form factors. However, this form-factor ratio can either be
measured experimentally, or eliminated by considering a double ratio
of amplitudes. In all cases, one is left only with a second-order
effect, $\sim 5\%$. We find twelve pairs of $B$ decays to which this
method can be applied. Depending on the pair, we estimate the total
theoretical error in relating the $B' \to M'_1 M'_2$ and $B^0 \to M_1
M_2$ amplitudes to be between 5\% and 15\%. The most promising decay
pairs are $\bd \to \pi^+ \pi^-$ and $B_u^+ \to K^0 \pi^+$, and $\bd
\to D^+ D^-$ and $\bd \to D_s^+ D^-$ or $B_u^+ \to D_s^+ {\bar
D}^0$. }


\preprint{UdeM-GPP-TH-04-119 \\ McGill 05/04}

\begin{document}

\newpage

\section{Introduction}

Within the standard model (SM), CP violation arises because of a
complex phase in the Cabibbo-Kobayashi-Maskawa (CKM) matrix. This
phase information can be encoded in the so-called unitarity triangle,
whose interior angles are known as $\alpha$, $\beta$ and $\gamma$
\cite{pdg}. The independent measurement of each of these CP angles
will overconstrain the unitarity triangle, thereby testing the SM.

One can perform further tests of the SM by measuring these angles in
many different ways. With this in mind, numerous methods, usually
involving $B$ decays, have been proposed for measuring the CP phases
\cite{CPreview}. In general, these techniques suffer from some degree
of theoretical error due to the hadronic uncertainty in $B$ decays.
Some methods are quite clean, i.e.\ they have little theoretical
uncertainty (e.g.\ the extraction of $\beta$ in $\bd(t) \to J/\psi
\ks$), while others have considerable hadronic uncertainty. Obviously,
the most interesting strategies are those which have as small a
theoretical error as possible.

One class of methods relies on flavor SU(3) symmetry \cite{su3}. Under
this symmetry, the quarks $u$, $d$ and $s$ all are placed in the same
multiplet. As a result, many particles are related by
SU(3)\footnote{Some methods refer to U-spin symmetry, which
interchanges $d$ and $s$ quarks. U-spin is a subgroup of the full
SU(3) symmetry.}, including $\pi$'s and $K$'s, $\bd$ and $\bs$ mesons,
and $D$ and $D_s$ mesons. Many techniques have been proposed to obtain
weak phase information from decays related by SU(3). Unfortunately,
all of these methods suffer from hadronic uncertainties due to the
breaking of the SU(3) symmetry, typically $O(m_s/\Lambda_{QCD}) \sim
25\%$. In order to obtain precise CP phase information from a given
method, this theoretical error must be reduced. For example, in some
techniques, the leading-order SU(3) breaking can be cast as the ratio
of (measured) decay constants, leaving a theoretical uncertainty at
the level of second-order SU(3) breaking.

Recently, we proposed two new methods, based on SU(3), for obtaining
CP phases. The measurements of $B^0_{d,s} \to K^{(*)} {\bar K}^{(*)}$
yield the angle $\alpha$ \cite{BKKbar}, while $\gamma$ can be
extracted from $\bd(t) \to D^{(*)+} D^{(*)-}$ and $\bd \to D_s^{(*)+}
D^{(*)-}$ \cite{BDDbar}. In both cases, we argued that the
leading-order SU(3) breaking is under control. In Ref.~\cite{BKKbar}
this is because a double ratio is used, so that the leading-order
SU(3)-breaking effect cancels. In Ref.~\cite{BDDbar}, SU(3) breaking
is due mainly to the ratio of decay constants, $f_{D_s}/f_D$, which is
known quite precisely in lattice gauge theory. (If one is reluctant to
use this input, one can instead use a double ratio, as in
Ref.~\cite{BKKbar}.) The upshot is that the remaining theoretical
error is at the level of a second-order effect, $\sim 5\%$.

The purpose of the present paper is to further explore these
techniques As we will show, they are essentially the same method.
Furthermore, one can apply this method to other pairs of $B$ decays --
the size of the theoretical error depends on the particular decays
considered.

The basic idea of the method is the following. The amplitude for a
$B^0 \to M_1 M_2$ decay with a significant $\btod$ penguin
contribution can be written $A_u V_{ub}^* V_{ud} + A_c V_{cb}^* V_{cd}
+ A_t V_{tb}^* V_{td} = (A_u - A_t) V_{ub}^* V_{ud} + (A_c - A_t)
V_{cb}^* V_{cd}$, where we have used the unitarity of the CKM matrix
to eliminate the $V_{tb}^* V_{td}$ term. It has been shown that one
cannot obtain clean CKM phase information from the measurement of such
a decay -- one always needs (at least) one piece of theoretical input
\cite{LSS}. Now consider a second decay $B' \to M'_1 M'_2$ which
receives a significant $\btos$ penguin contribution. The amplitude for
this decay can be written $(A'_u - A'_t) V_{ub}^* V_{us} + (A'_c -
A'_t) V_{cb}^* V_{cs} \approx (A'_c - A'_t) V_{cb}^* V_{cs}$. Here we
have used the fact that $|V_{ub}^* V_{us}| \ll |V_{cb}^* V_{cs}|$. We
now make the theoretical assumption that $(A'_c - A'_t)$ can be
related to $(A_c - A_t)$ by flavor SU(3). In this case, the
measurement of the branching ratio for $B' \to M'_1 M'_2$ will give us
the necessary input to extract CKM phase information from measurements
of the first decay. There are two sources of theoretical uncertainty
in such a method. First, $(A'_c - A'_t)$ and $(A_c - A_t)$ may not be
exactly equal in the SU(3) limit. And second, there will necessarily
be some SU(3) breaking in relating the two amplitudes. In order to
minimize this theoretical error, one must choose the two decays
carefully.

The paper is organized as follows. In Section 2, we describe the
method in considerably more detail. We show that, because of CKM
unitarity, one can extract either $\alpha$ or $\gamma$. The next two
sections are somewhat more technical. The reader who wishes to skip
these details can move directly to Section 5. In Section 3 we examine
which pairs of decays are useful for this method. We choose decay
pairs for which $(A'_c - A'_t)$ and $(A_c- A_t)$ are exactly related
in the SU(3) limit or are related by SU(3) provided certain small
amplitudes can be neglected. We find twelve such pairs. The next step
is to estimate the SU(3) breaking for each of these pairs. This is
done in Section 4 using large $N_c$ QCD and QCD factorization
\cite{BBNS}. We summarize and discuss our findings in Section 5. In
all cases, although the size of SU(3) breaking is about 25\%, we argue
that the theoretical uncertainty can be reduced to the level of a
second-order effect, $\sim 5\%$. In some cases, this comes about
because the leading-order SU(3) breaking is given by a (known) ratio
of decay constants. In the other cases, the SU(3) breaking involves an
unknown ratio of form factors. We show that, for many decay pairs,
this ratio can be measured. Failing this, one can reduce the SU(3)
theoretical uncertainty by using a double ratio of amplitudes.
Depending on the pair of $B$ decays chosen, and how one combines the
various theoretical uncertainties, the total theoretical error in
relating the $B' \to M'_1 M'_2$ and $B^0 \to M_1 M_2$ amplitudes is
between 5\% and 15\%. We find that our method is most promising for
three decay pairs: $\bd \to \pi^+ \pi^-$ and $B_u^+ \to K^0 \pi^+$,
for which data is already available, and $\bd \to D^+ D^-$ and $\bd
\to D_s^+ D^-$ \cite{BDDbar} or $B_u^+ \to D_s^+ {\bar D}^0$.  Using
the latest data, we show how $\gamma$ can be obtained from
measurements of the decay pair $\bd \to \pi^+ \pi^-$ and $B_u^+ \to
K^0 \pi^+$ in Section 6. (This is essentially an update of
Ref.~\cite{GroRosner}.)  We conclude in Section 7.

\newpage

\section{Extraction of CP Phases: General Case}

In this section, we describe the method for obtaining CP phase
information in as general terms as possible. We do not refer to
specific decays here; these are studied in the next section.

\subsection{Method I}

Consider a neutral $B^0 \to M_1 M_2$ decay involving a $\btod$ penguin
amplitude. $B^0$ can be either a $\bd$ or a $\bs$ meson, and $M_1$ and
$M_2$ are two mesons. (If both $M_1$ and $M_2$ are vector mesons, the
final state can be considered as a single helicity state of $M_1
M_2$.) The decay $B^0 \to M_1 M_2$ can be a pure penguin decay, or can
involve both tree and penguin contributions. In the latter case, it is
assumed that the penguin amplitude is not negligible. In addition, we
only consider final states $M_1 M_2$ accessible to both $B^0$ and
$\Bbar$ mesons. Thus, one expects indirect CP violation in such
decays.

The general amplitude for $B^0 \to M_1 M_2$ can be written
\bea
A(B^0 \to M_1 M_2) & = & A_u V_{ub}^* V_{ud} + A_c V_{cb}^* V_{cd} +
A_t V_{tb}^* V_{td} \nn\\
& = & (A_u - A_t) V_{ub}^* V_{ud} + (A_c - A_t) V_{cb}^* V_{cd} \nn\\
& \equiv & \Aut\ e^{i \gamma} e^{i \delta_{ut}} + \Act\ e^{i
  \delta_{ct}} ~,
\label{Bfamp}
\eea
where $\Aut \equiv |(A_u - A_t) V_{ub}^* V_{ud}|$, $\Act \equiv |(A_c
- A_t) V_{cb}^* V_{cd}|$, and we have explicitly written the strong
phases $\delta_{ut}$ and $\delta_{ct}$, as well as the weak phase
$\gamma$. In passing from the first line to the second, we have used
the unitarity of the CKM matrix, $V_{ub}^* V_{ud} + V_{cb}^* V_{cd}
+ V_{tb}^* V_{td} = 0$, to eliminate the $V_{tb}^* V_{td}$ term.

The amplitude ${\overline A}$ describing the conjugate decay $\Bbar
\to {\overline M}_1 {\overline M}_2$ can be obtained from
Eq.~(\ref{Bfamp}) by changing the sign of $\gamma$. Below we assume
that $M_1 M_2$ is self-conjugate, ${\overline M}_1 {\overline M}_2 =
M_1 M_2$. We consider the case in which ${\overline M}_1 {\overline
M}_2 \ne M_1 M_2$ in Sec.~2.3.

The time-dependent measurement of $B^0(t)\to M_1 M_2$ allows one to
obtain the three observables
\bea
B &\equiv & \frac{1}{2} \left( |A|^2 + |{\overline A}|^2 \right) = 
\Act^2 + \Aut^2 + 2 \Act \, \Aut \cos\delta \cos\gamma ~, \nn \\
a_{dir} &\equiv & \frac{1}{2} \left( |A|^2 - |{\overline A}|^2 \right)
= - 2 \Act \, \Aut \sin\delta \sin\gamma ~, \\
\aI &\equiv & {\rm Im}\left( e^{-2i \beta} A^* {\overline A} \right) =
-\Act^2 \sin 2\beta - 2 \Act \, \Aut \cos\delta \sin (2 \beta + \gamma)
\nn\\
& & \hskip2.6truein
- \Aut^2 \sin (2\beta + 2 \gamma)~, \nn
\eea
where ${\delta}\equiv {\delta}^{ut} - {\delta}^{ct}$. It is useful to
define a fourth observable:
\bea
\aR & \equiv & {\rm Re}\left( e^{-2i \beta} A^* {\overline A} \right) =
\Act^2 \cos 2\beta + 2 \Act \, \Aut \cos\delta \cos (2 \beta + \gamma)
\nn\\
& & \hskip2.6truein
+ \Aut^2 \cos (2\beta + 2 \gamma)~.
\eea
The quantity $\aR$ is not independent of the other three observables:
\beq
\aR^2 = B^2 - a_{dir}^2 - \aI^2 ~.
\label{aRdef}
\eeq
Thus, one can obtain $\aR$ from measurements of $B$, $a_{dir}$ and
$\aI$, up to a sign ambiguity.

The three independent observables depend on five theoretical
parameters: $\Aut$, $\Act$, $\delta$, $\beta$, $\gamma$. Therefore one
cannot obtain CP phase information from these measurements \cite{LSS}.
However, one can partially solve the equations to obtain
\beq
\Act^2 = { \aR \cos(2\beta + 2\gamma) - \aI \sin(2\beta + 2\gamma) - B
\over \cos 2\gamma - 1} ~.
\label{gammacond}
\eeq
Thus, assuming that $2\beta$ is known from the measurement of CP
violation in $\bd(t)\to J/\psi \ks$, we could obtain $\gamma$ if we
knew the value of $\Act$.

Consider now a decay $B' \to M'_1 M'_2$ involving a $\btos$
penguin. We refer to this as the ``partner process.'' This decay is
related by SU(3) symmetry to $B^0 \to M_1 M_2$. (In Sec.~3, for a
given $B^0 \to M_1 M_2$ decay, we present a variety of possibilities
for the partner process.) In this case, $B'$ ($\bd$, $\bs$ or $B_u^+$)
and $M'_1 M'_2$ depend on the choice of the partner process. The
amplitude for $B' \to M'_1 M'_2$ can be written
\bea
A(B' \to M'_1 M'_2) & = & A'_u V_{ub}^* V_{us} + A'_c
V_{cb}^* V_{cs} + A'_t V_{tb}^* V_{ts} \nn\\
& = & \Autp\ e^{i \gamma} e^{i \delta'_{ut}} + \Actp\ e^{i
\delta'_{ct}} \nn\\
& \approx & \Actp\ e^{i \delta'_{ct}} ~,
\label{Bfampprime}
\eea
where $\Autp \equiv |(A'_u - A'_t) V_{ub}^* V_{us}|$ and $\Actp \equiv
|(A'_c - A'_t) V_{cb}^* V_{cs}|$. In writing the last line, we have
taken $\Autp \ll \Actp$. (Note that $V_{ub}^* V_{us}$ is much smaller
than $V_{cb}^* V_{cs}$: $\left\vert {V_{ub}^* V_{us} / V_{cb}^*
V_{cs}} \right\vert \simeq 2\%$; we assume that $(A'_u - A'_t)$ is not
greatly enhanced relative to $(A'_c - A'_t)$, so that $\Autp \ll
\Actp$.) That is, {\it the partner process is assumed to be dominated
by a single amplitude.} This assumption contributes only a small
theoretical error, at the percent level. Thus, the measurement of the
rate for $B' \to M'_1 M'_2$ yields $\Actp$.

We now make the SU(3) assumption that
\beq
{\lambda \Actp \over \Act} = 1 ~,
\label{assumption}
\eeq
where $\lambda = 0.22$ is the Cabibbo angle. Combined with the
relation in Eq.~(\ref{gammacond}), this allows one to extract
$\gamma$.

The theoretical uncertainty in this method is essentially given by the
degree to which Eq.~(\ref{assumption}) is violated. This can occur in
two ways. First, even in the SU(3) limit, one might have $\lambda
\Actp/\Act \ne 1$. We will only consider pairs of decays for which
this error is small, at most $\sim 5\%$. Second, there are
SU(3)-breaking effects in Eq.~(\ref{assumption}). For a given set of
decays, the size of this error can be estimated -- we expect it to be
of first order in SU(3) breaking, $O(m_s/\Lambda_{QCD}) \sim 25\%$. As
we will see, for certain pairs of $B$ decays, this breaking can be
expressed in terms of a ratio of decay constants. If these decay
constants are known, then the leading-order SU(3) breaking is under
control, leaving an unknown second-order effect of $\sim 5\%$. In this
case, the assumption of Eq.~(\ref{assumption}) allows one to obtain
CKM phase information with a reasonably small theoretical error. (This
is the method described to obtain $\gamma$ from $\bd(t) \to D^{(*)+}
D^{(*)-}$ and $\bd \to D_s^{(*)+} D^{(*)-}$ \cite{BDDbar}.) In what
follows, we will refer to this as Method I.

There is an alternative, equivalent way to describe this method.
Consider again the amplitude for $B^0 \to M_1 M_2$
[Eq.~(\ref{Bfamp})]. If CKM unitarity is used to instead eliminate the
$V_{cb}^* V_{cd}$ term, one has
\beq
A(B^0 \to M_1 M_2) = \Auc\ e^{i \gamma} e^{i \delta_{uc}} + \Atc\
  e^{-i\beta} e^{i \delta_{tc}} ~,
\label{Bfamp2}
\eeq
where $\Auc \equiv |(A_u - A_c) V_{ub}^* V_{ud}|$, $\Atc \equiv |(A_t
- A_c) V_{tb}^* V_{td}|$. In this parametrization the three
independent observables measured in $B^0(t)\to M_1 M_2$ depend on four
theoretical parameters: $\Auc$, $\Atc$, $\Delta \equiv {\delta}^{uc} -
{\delta}^{tc}$, $\alpha$. It is therefore still not possible to obtain
CP phase information from these measurements. However, one can express
\beq
\Atc^2 = {\aR \cos 2\alpha + \aI \sin 2\alpha - B \over \cos 2\alpha -
  1} ~.
\label{alphacond}
\eeq
If we knew $\Atc$ we could extract $\alpha$.

Using the logic described above, we consider the partner decay $B' \to
M'_1 M'_2$. When the $V_{cb}^* V_{cd}$ term in Eq.~(\ref{Bfampprime})
is eliminated, one obtains
\beq
A(B' \to M'_1 M'_2) \approx \Atcp\ e^{i \delta'_{tc}} ~,
\eeq
where $\Atcp \equiv |(A'_t - A'_c) V_{tb}^* V_{ts}|$. The measurement
of the branching ratio for $B' \to M'_1 M'_2$ therefore yields
$\Atcp$. 

Now the SU(3) relation between $\Atc$ and $\Atcp$ is \cite{GroRosner}
\beq
{\lambda \Atcp \over \Atc} = \lambda \left\vert {V_{ts} \over V_{td}}
\right\vert = {\sin\alpha \over \sin\gamma} ~.
\label{assumption2}
\eeq
Writing $\gamma = \pi - \alpha - \beta$, and assuming that $\beta$ has
been measured, one can extract $\alpha$ from Eq.~(\ref{alphacond}).

Based on the above discussion, it appears that, depending on which
parametrization of the amplitudes is used, one can extract either
$\gamma$ or $\alpha$. However, these are equivalent. In both cases, we
assume that $\beta$ is known. Since $\alpha + \beta + \gamma = \pi$,
knowledge of one of $\alpha$ or $\gamma$ allows one to derive the
other angle. This simply reflects the fact that the three CP phases
are not independent. (In Ref.~\cite{LSS}, this is referred to as the
``CKM ambiguity.'')

\subsection{Method II}

One can remove the leading-order SU(3)-breaking effect as
follows. Consider a second decay $B^0 \to \tildeM_1 \tildeM_2$, where
$\tildeM_{1,2}$ are either excited states, or different helicity
states, of $M_{1,2}$. The amplitude for $B^0 \to \tildeM_1 \tildeM_2$
is given by an expression analogous to Eq.~(\ref{Bfamp}):
\beq
A(B^0 \to \tildeM_1 \tildeM_2) = \tildeAut\ e^{i \gamma} e^{i
  {\tilde\delta}_{ut}} + \tildeAct\ e^{i {\tilde\delta}_{ct}} ~.
\eeq
The time-dependent measurement of $B^0(t) \to \tildeM_1 \tildeM_2$
allows one to obtain $\tildeaR$, $\tildeaI$ and $\tildeB$, analogous
to the observables in $B^0(t) \to M_1 M_2$. We then have
\beq
{\Act^2 \over \tildeAct^2} = { \aR \cos(2\beta + 2\gamma) - \aI
\sin(2\beta + 2\gamma) - B \over \tildeaR \cos(2\beta + 2\gamma) -
\tildeaI \sin(2\beta + 2\gamma) - \tildeB} ~.
\label{gammacondprime}
\eeq
As before, given an independent measurement of $2\beta$, the knowledge
of $\Act/\tildeAct$ would allow us to obtain $\gamma$.

This information can be obtained by considering a second partner
process, $B' \to \tildeM'_1 \tildeM'_2$, where $\tildeM'_{1,2}$ are
either excited states, or different helicity states, of
$M'_{1,2}$. Analogous to Eq.~(\ref{Bfampprime}), we have
\beq
A(B' \to \tildeM'_1 \tildeM'_2) \approx \tildeActp\ e^{i
{\tilde\delta}'_{ct}} ~.
\eeq
The measurement of the rate for $B' \to \tildeM'_1 \tildeM'_2$ yields
$\tildeActp$. Now the assumption that
\beq
{\Actp / \tildeActp \over \Act / \tildeAct} = 1
\label{assumptionprime}
\eeq
provides the information necessary to obtain $\gamma$ from
Eq.~(\ref{gammacondprime}). 

Because we rely on a double ratio, we expect a significant
cancellation of the SU(3)-breaking effects. For example, this occurs
in decays where the leading-order SU(3) breaking is expressible in
terms of decay constants. In this case, the decay constants cancel in
Eq.~(\ref{assumptionprime}) for particular pairs of processes, leaving
only a second-order correction of $\sim 5\%$. In the more general
case, where SU(3) breaking involves also form factors, there is no
{\it proof} that the leading-order SU(3)-breaking effect cancels in
the double ratio. However, it is intuitively reasonable and this
cancellation, under certain conditions, can be demonstrated for
particular final states. In general significant cancellation of SU(3)
breaking effects in ratios of form factors are found in all explicit
calculations.

The theoretical error in Eq.~(\ref{assumptionprime}) ($\sim 5\%$) is
therefore considerably smaller than that of Eq.~(\ref{assumption})
($\sim 25\%$). For this reason it is often more advantageous to use
the method with the double ratio. (For specific decays, we provide
quantitative estimates of these theoretical uncertainties in Sec.~4.)
Henceforth, we will refer to this as Method II.

If one uses the parametrization of Eq.~(\ref{Bfamp2}), then the
relation analogous to Eq.~(\ref{gammacondprime}) is
\beq
{\Atc^2 \over \tildeAtc^2} = { \aR \cos 2\alpha + \aI \sin 2\alpha - B
\over \tildeaR \cos 2\alpha + \tildeaI \sin 2\alpha - \tildeB} ~.
\label{alphacondprime2}
\eeq
(Note that $\Act^2/\tildeAct^2 = \Atc^2/\tildeAtc^2$, so that
Eq.~(\ref{alphacondprime2}) is actually {\it identical} to
Eq.~(\ref{gammacondprime}).) In order to obtain the ratio on the
left-hand side, we once again consider the second partner decay, $B'
\to \tildeM'_1 \tildeM'_2$:
\beq
A(B' \to \tildeM'_1 \tildeM'_2) \approx \tildeAtcp\ e^{i
{\tilde\delta}'_{tc}} ~.
\eeq
The measurement of the rate for $B' \to \tildeM'_1 \tildeM'_2$ yields
$\tildeAtcp$. One then makes the assumption that
\beq
{\Atcp / \tildeAtcp \over \Atc / \tildeAtc} = 1 ~.
\label{assumptionprime2}
\eeq
In the case the factor of $\sin\alpha / \sin\gamma$ in
Eq.~(\ref{assumption2}) cancels between numerator and denominator.
Thus, Method II allows one to extract $\alpha$ from
Eq.~(\ref{alphacondprime2}) {\it without} assuming knowledge of
$\beta$. (This is the method described in Ref.~\cite{BKKbar}.)

\subsection{Method II$^{\prime}$}

Finally, we briefly discuss the case in which the final state in the
decay $B^0 \to M_1 M_2$ is not self-conjugate: ${\overline M}_1
{\overline M}_2 \ne M_1 M_2$. We describe the method for the
parametrization of Eq.~(\ref{Bfamp2}). As we will see below, the CP
phase $\alpha$ can be extracted. This is equivalent to the method in
which $\gamma$ is obtained, assuming that $\beta$ is known.

One now must consider separately the two decays $B^0(t)\to M_1 M_2$
and $B^0(t)\to {\overline M}_1 {\overline M}_2$:
\bea
A(B^0 \to M_1 M_2) & = & \Auc\ e^{i \gamma} e^{i \delta_{uc}} + \Atc\
  e^{-i\beta} e^{i \delta_{tc}} ~, \nn\\
A(B^0 \to {\overline M}_1 {\overline M}_2) & = & \tildeAut\ e^{i
  \gamma} e^{i {\tilde\delta}_{ut}} + \tildeAtc\ e^{-i\beta} e^{i
  {\tilde\delta}_{tc}} ~. \nn\\
\eea
As before, one can extract the following observables from the
time-dependent measurements of these decays: $B$, $a_{dir}$, $\aI$,
$\aR$, $\tildeB$, ${\tilde a}_{dir}$, $\tildeaI$ and $\tildeaR$. With
a bit of algebra, one can derive the following expression
\cite{BKKbar}:
\beq
W = X \tan 2\alpha + {\tildeAtc\over\Atc} \, {Y \over 2 \cos 2\alpha} -
{\Atc\over\tildeAtc} \, {Z \over 2 \cos 2\alpha} ~,
\label{alphacond2}
\eeq
where
\bea
W & \equiv & {1\over 2} (\aR - \tildeaR) ~, \nn\\ 
X & \equiv & {1\over 2} (-\aI + \tildeaI) ~, \nn\\ 
Y & \equiv & {1\over 2} (-B -a_{dir} + \tildeB\ - {\tilde a}_{dir}) ~,
\nn\\
Z & \equiv & {1\over 2} (B -a_{dir} - \tildeB\ - {\tilde a}_{dir}) ~.
\eea
If we knew $\Atc / \tildeAtc$ we could extract $\alpha$.

This information can be obtained from the partner processes $B' \to
M'_1 M'_2$ and $B' \to {\overline M}'_1 {\overline M}'_2$ which are
each dominated by a single decay amplitude. The measurement of the
rates for these decays allows one to extract $\Atcp$ and
$\tildeAtcp$. We assume that
\beq
{\Atcp / \tildeAtcp \over \Atc / \tildeAtc} = 1 ~.
\eeq
This provides the theoretical input necessary to obtain $\alpha$ from
Eq.~(\ref{alphacond2}). As before, the leading-order SU(3)-breaking
effect cancels in the double ratio, so that the net theoretical error
is a second-order effect.

\vskip3truemm
We have therefore shown that it is possible to extract CP phase
information from time-dependent measurements of the decays $B^0(t) \to
M_1 M_2$ and $B^0(t) \to \tildeM_1 \tildeM_2$, along with rate
measurements of their SU(3)-related partner processes $B' \to M'_1
M'_2$ and $B' \to \tildeM'_1 \tildeM'_2$. The above discussion has
been completely general; in the next section we turn to an examination
of the specific decays to which this method can be applied.

\section{Specific Decays}

Here we examine the decays to which the method described in the
previous section can be applied. The first step is to find neutral
$B^0 \to M_1 M_2$ decays involving a $\btod$ penguin amplitude, with
the condition that both $B^0$ and $\Bbar$ mesons can decay to $M_1
M_2$. Such decays are straightforward to tabulate. They are:
\bea
& \bd \to D^+ D^- ~,~ \pi^+ \pi^- ~,~ \pi^0 \pi^0 ~,~ K^0 {\bar K}^0
~, & \nn\\
& \bs \to  {\bar K}^0 \pi^0 ~,~ \eta_s  {\bar K}^0 ~. &
\label{BMMdecays}
\eea
In all cases the final state is written in terms of pseudoscalars
(P's) only. However, it is understood that either or both of the
final-state particles can be vector mesons (V's). In the case of two
vector mesons, there are three helicity states; each of these can be
considered as a separate final state.

Recall that we require either that the final state be a CP eigenstate
(Methods I and II), or that the $B^0$ be able to decay to both $M_1
M_2$ and ${\overline M}_1 {\overline M}_2$ (Method II$^\prime$). As
written, the $\bs$ decays above do not satisfy either of these
conditions because of the presence of the `${\bar K}^0$' in the final
state. (As noted above, this is to be understood as either a P or a
V.) In order to apply our methods to these decays, this particle must
be a $\ks$ if it is a pseudoscalar. If the final-state particle is a
${\bar K}^{*0}$, it must be detected via its decay to $\ks \pi^0$.

In Eq.~(\ref{BMMdecays}), $\eta_s$ corresponds to a pure $\ssbar$
quark pair. In practice, if the $\ssbar$ hadronizes as a pseudoscalar,
it will be found as either an $\eta$ or $\eta'$ meson, both of which
also have significant $(u \bar u)$ and $(d \bar d)$ components
\cite{eta}. As a result, in decays involving an $\eta$ or $\eta'$,
$B^0 \to M_1 M_2$ and $B' \to M'_1 M'_2$ are not really related by
SU(3), and our method does not apply. It is therefore better to
consider the vector meson $\phi$ which is a pure $\ssbar$ quark state
to a very good approximation.

The second step is to find partner processes $B' \to M'_1 M'_2$,
related to $B^0 \to M_1 M_2$ by SU(3) symmetry, which involve a
$\btos$ penguin amplitude. This is more complicated, as there are
several possibilities.

Consider first the decay $\bd \to D^+ D^-$, which receives
contributions from a ${\bar b} \to {\bar d} c {\bar c}$ penguin
amplitude. Under SU(3), the $d$, $s$ and $u$ quarks are treated on an
equal footing. Therefore there are three possible candidate partner
processes for this decay. The $\btod$ penguin amplitude is changed to
a $\btos$ amplitude, and one considers three different flavors for the
spectator quark:
\beq
\bs \to D_s^+ D_s^- ~,~ \bd \to D_s^+ D^- ~,~ B_u^+ \to D_s^+ {\bar
D}^0 ~.
\eeq

The remaining five decays in Eq.~(\ref{BMMdecays}) all involve the
decay of a ${\bar b}$ quark into light quarks. Since all three light
quarks are equivalent under SU(3), the potential partner processes are
given by a $\bd$, $\bs$ or $B_u^+$ meson decaying via a $\btos$
penguin diagram, with the gluon (or $Z$) turning into a ${\bar u} u$,
${\bar d} d$ or ${\bar s} s$ quark pair. That is, the candidate
partner processes for all five decays are
\bea
\bd & \to & K^+ \pi^- ~,~ K^0 \pi^0 ~,~ \eta_s K^0 ~, \nn\\
\bs & \to & K^+ K^- ~,~ K^0 {\bar K}^0 ~,~ \eta_s \eta_s ~, \nn\\
B_u^+ & \to & K^+ \pi^0 ~,~ K^0 \pi^+ ~,~ \eta_s K^+ ~.
\eea

Now, recall that one of the requirements of the method is that the
partner process be dominated by a single amplitude. Unfortunately, the
decays $\bd \to K^+ \pi^-$, $\bs \to K^+ K^-$ and $B_u^+ \to K^+
\pi^0$ above receive significant contributions from both tree and
$\btos$ penguin amplitudes. Thus, they are not dominated by a single
amplitude, and do not satisfy the above requirement. Therefore, they
cannot be used as partner processes in our method.

(On the other hand, we note that Ref.~\cite{Fleischer1} uses U-spin
symmetry to relate $\bd \to \pi^+ \pi^-$ to $\bs \to K^+ K^-$ (or $\bd
\to K^+ \pi^-$, if exchange-type diagrams are neglected). Here it is
assumed that direct CP violation is measured in the partner process.
With the assumption of U-spin symmetry, one can extract $\gamma$. It
has been argued that the SU(3)-breaking corrections in this approach
may be sizeable \cite{BsKKSU3break}. However, this occurs if
annihilation contributions are large, which is not the naive
expectation. We discuss the SU(3)-breaking effects in such
contributions in Sec.~4.7.)

Most pairs of decays ($B^0 \to M_1 M_2$ and its partner process) have
not been studied, but there are a few exceptions. As already noted, we
have examined $\bd\to K^0{\bar K}^0$ and $\bs\to K^0{\bar K}^0$
\cite{BKKbar}, as well as $\bd\to D^+ D^-$ and $\bd\to D_s^+ D^-$
\cite{BDDbar}. In addition, Fleischer has noted that one can get
$\gamma$ from $\bd\to D^+ D^-$ and $\bs\to D_s^+ D_s^-$ decays
\cite{Fleischer2}. In this paper one keeps both contributing
amplitudes to $\bs\to D_s^+ D_s^-$, and assumes that their (small)
interference is measured. By using U-spin to relate $\bd\to D^+ D^-$
and $\bs\to D_s^+ D_s^-$, one can obtain $\gamma$. However, if one
neglects the small $V_{ub}^* V_{us}$ amplitude in $\bs\to D_s^+
D_s^-$, one essentially reproduces the method outlined in Sec.~2.
Finally, the pair $\bd \to \pi^+ \pi^-$ and $B_u^+ \to K^0 \pi^+$ has
been studied in Ref.~\cite{GroRosner}.

In the remainder of this section, we will examine all $B^0 \to M_1
M_2$ decays, along with their potential partner processes, to see
which can be used to obtain CP phase information with our method. At
this stage the goal is simply to find pairs of processes in which the
amplitudes $\Act$ and $\lambda\Actp$ (or, equivalently, $\Atc$ and
$\lambda\Atcp$) are equal in the SU(3) limit; SU(3) breaking will be
studied in the next section.

Within SU(3), all $B^0 \to M_1 M_2$ decays can be expressed in terms
of a small number of matrix elements. This is equivalent to a
description in terms of diagrams \cite{firstSU3,EWPs}. Including
electroweak penguin contributions (EWP's), there are eight main
contributing diagrams: (1) a color-favored tree amplitude $T$, (2) a
color-suppressed tree amplitude $C$, (3) a gluonic penguin amplitude
$P$, (4) an exchange amplitude $E$, (5) an annihilation amplitude $A$,
(6) a penguin annihilation amplitude $PA$, (7) a color-favored
electroweak penguin amplitude $\pew$, and (8) a color-suppressed
electroweak penguin amplitude $\pewc$. For ${\bar b} \to {\bar d} q
{\bar q}$ and ${\bar b} \to {\bar s} q {\bar q}$ ($q=u,~d,~s$), we
write the diagrams with no primes and primes, respectively. For ${\bar
b} \to {\bar d} c {\bar c}$ and ${\bar b} \to {\bar s} c {\bar c}$, we
write the diagrams with tildes and tildes plus primes, respectively.
We will express all amplitudes in terms of these diagrams.

Pairs of decays whose amplitudes are the same, except for primes, are
equal in the SU(3) limit. Any difference in the amplitudes for a pair
of decays will lead to a theoretical error. To estimate this error, a
useful rule of thumb is the approximate sizes of the various diagrams
\cite{EWPs}. For ${\bar b} \to {\bar d} q {\bar q}$ processes
($q=u,~d,~s$), they are
\bea
1 & : & |T|, \nn \\
{\cal O}({\bar\lambda}) & : & |C|,~|P|, \nn \\
{\cal O}({\bar\lambda}^2) & : & |E|,~|A|,~|\pew| \nn \\
{\cal O}({\bar\lambda}^3) & : & |PA|,~|\pewc|,
\label{buudhierarchy}
\eea
while for ${\bar b} \to {\bar s} q {\bar q}$ transitions
($q=u,~d,~s$), one has
\bea
1 & : & |P'|, \nn \\
{\cal O}({\bar\lambda}) & : & |T'|,~|\pew'|\nn \\
{\cal O}({\bar\lambda}^2) & : & |C'|,~|PA'|,~|\pewcp| \nn \\
{\cal O}({\bar\lambda}^3) & : & |E'|,~|A'|~.
\label{buushierarchy}
\eea
In the above, ${\bar\lambda} \sim 20\%$. There are a variety of
sources for these suppressions: (1) CKM matrix elements: e.g.\ $T'
\sim V_{ub}^* V_{us} \sim {\bar\lambda}^4$, $P' \sim V_{cb}^* V_{cs}
\sim {\bar\lambda}^2$, (2) loop factors: e.g.\ all penguin and
electroweak penguin diagrams arise at loop level, (3) color
suppression: e.g.\ $C$ is smaller than $T$ by a factor of about
${\bar\lambda}$, (4) $m_t$: although electroweak penguin amplitudes
are smaller than penguin contributions by a factor of
$\alpha/\alpha_s$, their numerical importance is enhanced by a factor
of $m_t^2/M_Z^2$ \cite{GSW}, (5) exchange- and annihilation-type
diagrams are suppressed by $f_B/m_B \sim {\bar\lambda}^2$. Putting all
these suppression factors together, most of which are reflected in the
Wilson coefficients for the various operators, one arrives at the
hierarchies above. We stress that we have implicitly assumed that the
ratios of matrix elements do not differ significantly from unity. Thus
these sizes are to be taken as rough estimates only.

Note that, for both $\btod$ and $\btos$ transitions, the exchange and
annihilation contributions are expected to be quite small. However, in
some approaches to hadronic $B$ decays, such amplitudes may be
chirally enhanced if there are pseudoscalars in the final state
\cite{BBNS,PQCD}. On the other hand, such chiral enhancements are not
present for vector-vector final states. Ultimately, the size of
exchange and annihilation diagrams is an experimental question, and
can be tested by the measurement of decays such as $\bd \to D_s^+
D_s^-$ and $\bd\to K^+ K^-$.

The amplitude for $\bd \to D^+ D^-$ is given by
\beq
A(\bd \to D^+ D^-) = \tildeT\ + \tildeE\ + \tildeP\ + \tildePA\ +
{2\over 3} \tildepewc\ ~,
\label{BDDamp}
\eeq
while those of the potential partner processes are
\bea
A(\bs \to D_s^+ D_s^-) & = & \tildeT' + \tildeE' + \tildeP' +
\tildePA' + {2\over 3} \tildepewcp\ ~, \nn\\
A(\bd \to D_s^+ D^-) & = & \tildeT' + \tildeP' + {2\over 3}
\tildepewcp\ ~, \nn\\
A(B_u^+ \to D_s^+ {\bar D}^0) & = & \tildeT' + \tildeP' + \tildeA' +
{2\over 3} \tildepewcp\ ~, \nn\\
\label{BDDpartneramp}
\eea

The amplitudes for the remaining five $B^0 \to M_1 M_2$ decays in
Eq.~(\ref{BMMdecays}) are given by
\bea
A(\bd \to \pi^+ \pi^-) & = & - T - P - E - PA - {2 \over 3} \pewc ~,
\nn\\
A(\bd \to K^0 {\bar K}^0) & = & P + PA - {1 \over 3} \pewc ~, \nn\\
\sqrt{2} A(\bd \to \pi^0 \pi^0) & = & P - C + E + PA - \pew - {1\over
3} \pewc ~, \nn\\
\sqrt{2} A(\bs \to {\bar K}^0 \pi^0) & = & P - C - \pew - {1\over 3}
\pewc ~, \nn\\
A(\bs \to \eta_s {\bar K}^0) & = & P - {1\over 3} \pew - {1\over
3} \pewc ~.
\label{BMMamps}
\eea
The amplitudes for the six potential partner processes are
\bea
\sqrt{2} A(\bd \to K^0 \pi^0) & = & P' - C' - \pew' - {1 \over 3}
\pewcp\ ~, \nn\\
A(\bd \to \eta_s K^0) & = & P' - {1 \over 3} \pew' - {1 \over 3}
\pewcp\ ~, \nn\\
A(\bs \to K^0 {\bar K}^0) & = & P' + PA' - {1 \over 3} \pewcp\ ~, \nn\\
A(\bs \to \eta_s \eta_s) & = & P' + PA' - {1 \over 3} \pew' - {1
\over 3} \pewcp\ ~, \nn\\
A(B_u^+ \to K^0 \pi^+) & = & P' + A' - {1\over 3} \pewcp\ ~, \nn\\
A(B_u^+ \to \eta_s K^+) & = & P' + A' - {1\over 3} \pew' - {1\over 3}
\pewcp\ ~.
\label{partneramps}
\eea

For each of the decays in Eq.~(\ref{BMMamps}), we wish to find the
process(es) in Eq.~(\ref{partneramps}) whose amplitude is equal in the
SU(3) limit. However, recall that we are comparing $\Act$ and
$\lambda\Actp$, i.e.\ the pieces of the amplitudes proportional to
$V_{cb}^* V_{cd}$ ($B^0 \to M_1 M_2$) or $V_{cb}^* V_{cs}$ (partner
process). Several pieces in the amplitudes of Eq.~(\ref{BMMamps}) are
proportional to $V_{ub}^* V_{ud}$: the $T$ and $E$ amplitudes in $\bd
\to \pi^+ \pi^-$, the $C$ and $E$ amplitudes in $\bd \to \pi^0 \pi^0$,
and the $C$ amplitude in $\bs \to {\bar K}^0 \pi^0$. In comparing the
amplitudes of Eqs.~(\ref{BMMamps}) and (\ref{partneramps}), these
pieces are unimportant.

Finally, we have noted that, for both the decay and its partner
process, either or both of the final-state mesons can be a
pseudoscalar or a vector. However, one has to be careful if PV states
are used. Within QCD factorization \cite{BBNS}, which we use to
calculate SU(3) breaking in Sec.~4, one of the final-state mesons
comes directly from the decay of the $B$, while the other is produced
from the vacuum. If the ``vacuum meson'' for the decay is a P, but is
a V for the partner process (or vice-versa), then the two decays are
{\it not} related by SU(3). The reason is that the two processes
receive contributions from different operators. For example, those
operators responsible for chiral enhancement affect the production of
a P from the vacuum, but not a V. Thus, although the decay and its
partner process are related by SU(3) at the quark level, they are not
related at the meson level.

We consider each of the $B^0 \to M_1 M_2$ decays in turn, labeling
each of the subsections by the decay in Eq.~(\ref{BMMdecays}).

\subsection{$\bd \to D^+ D^-$}

$\bd \to D^+ D^-$ is unique among $B^0 \to M_1 M_2$ decays in that its
largest decay component is a ${\bar b} \to c {\bar c} d$ tree
amplitude $\tildeT$. Referring to Eqs.~(\ref{BDDamp}) and
(\ref{BDDpartneramp}), we note that the amplitude for $\bs \to D_s^+
D_s^-$ is equal to that of $\bd \to D^+ D^-$ in the SU(3) limit. As
for $\bd \to D_s^+ D^-$, its amplitude differs from that of $\bd \to
D^+ D^-$ only by the diagrams $\tildeE$ and $\tildePA$. A theoretical
error is incurred in neglecting these contributions due to the fact
that $|\tildeE/\tildeT| \sim 5\%$ and $|\tildePA/\tildeT| \sim
1\%$. The decay $B_u^+ \to D_s^+ {\bar D}^0$ is similar to $\bd \to
D_s^+ D^-$, except that one has to neglect in addition the $\tildeA'$
piece. However, since it is proportional to $V_{ub}^* V_{us}$,
$\tildeA'$ is expected to be tiny. Thus, all decays can be used as
partner processes within our method, though there is a theoretical
error of $\sim 5\%$ coming from the amplitudes if $\bd \to D_s^+ D^-$
\cite{BDDbar} or $B_u^+ \to D_s^+ {\bar D}^0$ are used.

\subsection{$\bd \to \pi^+ \pi^-$}

None of the amplitudes in Eq.~(\ref{partneramps}) is a perfect match
for the amplitude of $\bd \to \pi^+ \pi^-$ [Eq.~(\ref{BMMamps})]. Of
the six potential partner processes, four receive contributions from
$\pew'$, which is not small: $|\pew'/P'| \sim 20\%$
[Eq.~(\ref{buushierarchy})]. Thus, if any of these decays is used as
the partner process, the theoretical error will be at least $\sim
20\%$. Note that there is a possible loophole here. The EWP
contribution to three of the four potential partner processes is
actually $(1/3)\pew'$. In this case $(1/3)|\pew'/P'| \sim 7\%$, which
is not that large. Thus, these decays could perhaps be used as partner
processes. However, this could easily be spoiled by a large ratio of
matrix elements, and so we do not consider these decays as partner
processes. However, the remaining two decays can be used as partner
processes. In the SU(3) limit the difference in the amplitudes of $\bs
\to K^0 {\bar K}^0$ and $\bd \to \pi^+ \pi^-$ is at the level of
$|\pewcp/P'| \sim 5\%$. $B_u^+ \to K^0 \pi^+$ is slightly worse: there
are additional differences of $|A'/P'| \sim 1\%$ and $|PA/P| \sim
5\%$. Thus, both of these decays can be used as partner processes,
incurring a small theoretical error due to amplitude differences in
the SU(3) limit.

\subsection{$\bd \to K^0 {\bar K}^0$}

The decay $\bd \to K^0 {\bar K}^0$ has no $\pew$ component. Thus, in
searching for partner processes to $\bd \to K^0 {\bar K}^0$, we can
exclude the four processes in Eq.~(\ref{partneramps}) which receive
contributions from $\pew'$ (see, however, the discussion in $\bd \to
\pi^+ \pi^-$ above). Of the remaining two decays, only one has an
amplitude which is equal to that of $\bd \to K^0 {\bar K}^0$ in the
SU(3) limit: $\bs \to K^0 {\bar K}^0$. The other, $B_u^+ \to K^0
\pi^+$, is a reasonable match, but not perfect. If it is used as a
partner process, there is a theoretical error due to amplitude
differences at the level of $|A'/P'| \sim 1\%$ and $|PA/P| \sim 5\%$.

\subsection{$\bd \to \pi^0 \pi^0$}

Although the final state is written as $\pi^0\pi^0$, it is understood
that in this case the vector-vector state $\rho^0\rho^0$ must be used.
The method requires the time-dependent measurement of the decay $B^0
\to M_1 M_2$. However, this is virtually impossible for $\bd \to \pi^0
\pi^0$ since it is very difficult to find the vertex of a $\pi^0$.

The amplitude for the decay $\bd \to \pi^0 \pi^0$ has a $\pew$
component. Five of the processes in Eq.~(\ref{partneramps}) receive
different contributions from $\pew'$, and can therefore be excluded as
partner processes. Only $\bd \to K^0 \pi^0$ is a good match: the
amplitudes proportional to $V_{cb}^*V_{cq}$ ($q=d,s$) are equal (apart
from primes), except for the $PA$ piece in $\bd \to \pi^0 \pi^0$.
There is also a $C'$ piece in $\bd \to K^0 \pi^0$ which is
proportional to $V_{ub}^*V_{us}$. However, both $PA$ and $C'$ are
expected to be small: referring to Eqs.~(\ref{buudhierarchy}) and
(\ref{buushierarchy}), we see that both $|PA/P|$ and $|C'/P'|$ are
expected to be $\sim 5\%$. Because $C'$ is very small, the requirement
that the partner process be dominated by a single decay amplitude is
satisfied.

Thus, we can apply our method to the decays $\bd \to \pi^0 \pi^0$ and
$\bd \to K^0 \pi^0$, incurring a theoretical uncertainty of $\sim
10\%$ due to the difference of the amplitudes. Since we know that the
VV final state must be used for this particular pair of decays, from
now on we will refer to them as $\bd\to\rho^0\rho^0$ and $\bd\to
K^{*0}\rho^0$.

\subsection{$\bs \to {\bar K}^0 \pi^0$}

Like $\bd \to \pi^0 \pi^0$, the amplitude for $\bs \to {\bar K}^0
\pi^0$ has a $\pew$ component. Thus, the only decay in
Eq.~(\ref{partneramps}) that can be considered as a partner process is
$\bd \to K^0 \pi^0$. For the other decays, the theoretical error due
to the amplitude differences is at the level of $|\pew/P| \sim
|\pew'/P'| \sim 20\%$. For $\bd \to K^0 \pi^0$, the theoretical error
is only $\sim 5\%$ due to the neglect of the $C'$ amplitude. 

As in $\bd \to \pi^0 \pi^0$ above, the final-state `$\pi^0$' must be a
$\rho^0$. The `${\bar K}^0$' can in principle be either a pseudoscalar
or a vector. However, if it is a P, then, as explained earlier, the
decay $\bs \to {\bar K}^0 \rho^0$ is {\it not} related by SU(3) to
$\bd \to K^0 \rho^0$. The reason is that, in $\bs \to {\bar K}^0
\rho^0$, the $\rho^0$ is produced from the vacuum, whereas in $\bd \to
K^0 \rho^0$ it is the $K^0$. Since this particle is a pseudoscalar
meson in one case, but a vector meson in the other, different
operators are involved and the two decays are not related by SU(3).
Thus, the vector-vector final states must be used for both the decay
and partner process. Henceforth we refer to this pair of decays as
$\bs \to {\bar K}^{*0} \rho^0$ and $\bd \to K^{*0}\rho^0$. Recall that
we require that both $\bs$ and $\bsbar$ decay to the same final state.
In order for this to be possible, the ${\bar K}^{*0}$ must be detected
via its decay to $\ks \pi^0$ (as in the measurement of $\sin 2\beta$
via $\bd(t) \to J/\psi K^*$).

\subsection{$\bs \to \eta_s {\bar K}^0$}

In this case, there are three decays which can be taken as partner
processes. The amplitude for $\bd \to \eta_s K^0$ is equal to that of
$\bs \to \eta_s {\bar K}^0$ in the SU(3) limit. The amplitude for
$B_u^+ \to \eta_s K^+$ differs only by $|A'/P'| \sim 1\%$. Finally,
since $|PA'/P'|$ is very small [Eq.~(\ref{buushierarchy})], $\bs \to
\eta_s \eta_s$ can also be taken as the partner process, though there
is a small ($\sim 5\%$) theoretical error coming from the amplitudes.
The other decays have theoretical errors of about $|\pew/P| \sim
|\pew'/P'| \sim 20\%$ due to differences in the amplitudes.

Thus, $\bd \to \eta_s K^0$, $B_u^+ \to \eta_s K^+$ and $\bs \to \eta_s
\eta_s$ can all be taken as partner processes to $\bs \to \eta_s {\bar
K}^0$. As discussed previously, the $(s \bar s)$ state is best viewed
as a vector $\phi$ meson in our method. In addition, as was the case
for $\bs \to {\bar K}^0 \pi^0$, the final-state $K$-meson must be a
${\bar K}^{*0}$, detected via its decay to $\ks \pi^0$. {}From now on
we will therefore refer to these decays via their VV final states.

\newpage
\section{SU(3) Breaking}

We have found twelve pairs of decays to which our method can be
applied. They are
\begin{enumerate}

\item $\bd \to D^+ D^-$ and $\bs \to D_s^+ D_s^-$, $\bd \to D_s^+
  D^-$, or $B_u^+ \to D_s^+ {\bar D}^0$;

\item $\bd \to \pi^+ \pi^-$ and $\bs \to K^0 {\bar K}^0$ or $B_u^+ \to
  K^0 \pi^+$;

\item $\bd \to K^0 {\bar K}^0$ and $\bs \to K^0 {\bar K}^0$ or $B_u^+
  \to K^0 \pi^+$;

\item $\bd\to\rho^0\rho^0$ and $\bd\to K^{*0}\rho^0$;

\item $\bs \to {\bar K}^{*0} \rho^0$ and $\bd \to K^{*0}\rho^0$;

\item $\bs \to \phi {\bar K}^{*0}$ and $\bd\to\phi K^{*0}$, $B_u^+ \to
\phi K^{*+}$, or $\bs\to\phi\phi$.

\end{enumerate}
In all cases there is a theoretical error at the percent level due to
the neglect of the $V_{ub}^* V_{us}$ term in the partner process.
Some pairs of decays have an additional theoretical error due to
neglecting certain diagrams. For all of the above decay pairs, this
error is at the level of ${\bar\lambda}^2$ or less
[Eqs.~(\ref{buudhierarchy}) and (\ref{buushierarchy})].

The main source of theoretical error is SU(3) breaking. In the
following subsections, we examine the size of SU(3) breaking in each
of the six classes above.

Throughout we ignore SU(3) breaking in annihilation and exchange
amplitudes. There are two reasons for this. First, in most cases, the
partner process is related to the principal decay only if
annihilation/exchange diagrams are neglected. In this situation, SU(3)
breaking in such contributions is irrelevant. Second, even if the
decay pairs are perfectly related in the SU(3) limit, including
annihilation/exchange diagrams, there are no reliable theoretical
methods for estimating the size of such contributions. As a result,
the size of SU(3) breaking is also unreliable. Thus, our estimates of
SU(3) breaking exclude annihilation and exchange amplitudes. However,
we discuss these in greater detail in Sec.~4.7.

\subsection{$\bd \to D^+ D^-$}

There are three possible partner processes to $\bd \to D^+ D^-$: $\bs
\to D_s^+ D_s^-$, $\bd \to D_s^+ D^-$ and $B_u^+ \to D_s^+ {\bar
D}^0$. We refer to the three pairs of decays, in this order, as pair
``{\it a}'', ``{\it b}'' and ``{\it c}''. All amplitudes are dominated
by the tree contributions. The penguin and electroweak penguin
diagrams are smaller than the tree diagrams: the ratios
$\tildeP/\tildeT$, etc.\ are $\sim 25\%$ or smaller. Thus, SU(3)
breaking originates mainly in the ratio $\tildeT'/\tildeT$ in
Eq.~\ref{assumption} --- the SU(3)-breaking contribution from the
penguin amplitudes is of higher order. Factorization has been used to
study $B \to D^{(*)} {\bar D}^{(*)}$ decays \cite{luorosner}, and it
has been found that experiments are consistent with the factorization
predictions. We will therefore use factorization to estimate the SU(3)
breaking in the various decay pairs.

We begin by calculating $\tildeT$ and $\tildeT'$ within
factorization. These are generated by two terms in the effective
Hamiltonian \cite{BuraseffH}:
\beq
H_{tree}^q = {G_F \over \sqrt{2}} \left[ V_{ub}V^*_{uq}(c_1 O_1^q + c_2
O_2^q) \right] ~,
\eeq
where $q$ can be either a $d$ quark ($\bd \to D^{+}D^{-}$) or an $s$
quark (partner process). The operators $O_i^q$ are defined as
\beq
O_1^q = {\bar q}_\alpha \gamma_\mu L u_\beta \, {\bar u}_\beta
\gamma^\mu L b_\alpha ~~,~~ O_2^q = {\bar q} \gamma_\mu L u \, {\bar
u} \gamma^\mu L b ~,
\eeq
where $L = 1 - \gamma_5$, and $c_1$ and $c_2$ are Wilson
coefficients. Within factorization, the various tree amplitudes are
given by
\bea
|\tildeT( \bd \to D^+ D^-)| & = & \frac{G_F}{\sqrt 2} |V_{cb}V_{cd}^*|
\left( {c_1 \over N_c} + c_2 \right) f_{D} {F_0^{\bd \to D}(m_D^2)}
(m_B^2-m_D^2) ~, \nn\\
|\tildeT'( \bs \to D_s^+ D_s^-)| & = & {\frac{G_F}{\sqrt 2}}
|V_{cb}V_{cs}^*| \left( {c_1 \over N_c} + c_2 \right) f_{D_s} F_0^{\bs
\to D_s}(m_{D_s}^2) (m_{B_s}^2-m_{D_s}^2) ~, \nn\\
|\tildeT'( \bd \to D_s^+ D^-)| & = & {\frac{G_F}{\sqrt 2}}
|V_{cb}V_{cs}^*| \left( {c_1 \over N_c} + c_2 \right) f_{D_s} F_0^{\bd
\to D}(m_{D_s}^2) (m_B^2-m_D^2) ~, \nn\\
|\tildeT'( B_u^+ \to D_s^+ {\bar D}^0)| & = & {\frac{G_F}{\sqrt 2}}
|V_{cb}V_{cs}^*| \left( {c_1 \over N_c} + c_2 \right) f_{D_s} F_0^{\bd
\to D}(m_{D_s}^2) (m_B^2-m_D^2) ~,
\eea
where $N_c$ represents the number of colors. The form factor $F_0^{B
\to M}$ for a $B \to M$ transition is defined through \cite{BSW}
\bea
\langle M(p_K)|\bar{q} \gamma_\mu (1-\gamma_5) b | B(p_B) \rangle &=&
\left[ (p_B+ p_M)_\mu - \frac{m_B^2-m_M^2}{q^2} q_\mu \right] F_1^{ B
\to M}(q^2)\nn\\ 
& + & \frac{m_B^2-m_M^2}{q^2} q_\mu F_0^{ B \to M}(q^2) ~,
\label{formfactor}
\eea
where $q = p_B -p_M $. 

The size of SU(3)-breaking for the three decays pairs is given by the
deviation from unity of
\bea
\left( {\lambda \tildeT' \over \tildeT} \right)^a & = & {f_{D_s}
F_0^{\bs \to D_s}(m_{D_s}^2) (m_{B_s}^2-m_{D_s}^2) \over f_{D} {F_0^{B
\to D}(m_D^2)} (m_B^2-m_D^2)} ~, \nn\\
\left( {\lambda \tildeT' \over \tildeT} \right)^b & = & {f_{D_s}
F_0^{\bd \to D}(m_{D_s}^2) \over f_{D} {F_0^{\bd \to D}(m_D^2)}} \approx
{f_{D_s} \over f_D} ~, \nn\\
\left( {\lambda \tildeT' \over \tildeT} \right)^c & = & {f_{D_s}
F_0^{\bd \to D}(m_{D_s}^2) \over f_{D} {F_0^{\bd \to D}(m_D^2)}} \approx
{f_{D_s} \over f_D} ~.
\label{su3BDD}
\eea
Note that the form factor $F^{\bd \to D}_0$ can be related to the
Isgur-Wise function in the heavy quark limit. Since $F^{\bd \to D}_0$
is smooth, there is negligible error in setting $F^{\bd \to
D}_0(m_{D_s}^2) = F^{\bd \to D}_0(m_D^2)$. Thus, for pairs ``{\it b}''
and ``{\it c}'', SU(3) breaking is due almost entirely to the
difference between the $D$ and $D_s$ decay constants within
factorization.

The situation is worse for pair ``{\it a}''. Although the form factors
$F^{B_s \to D_s}_0(m_{D_s}^2)$ and $F^{\bd \to D}_0(m_{D}^2)$ are
equal in the SU(3) limit, the SU(3) breaking can be calculated in
chiral perturbation theory and may not be small \cite{su3isgur}. Thus,
it may be better to use $\bd \to D_s^+ D^-$ or $B_u^+ \to D_s^+ {\bar
D}^0$ as the partner process. Although there is a theoretical error,
not present in $\bs \to D_s^+ D_s^-$, due to neglecting small
amplitudes, this may be smaller than the error due to SU(3) breaking
in $F^{B_s \to D_s}_0(m_{D_s}^2)$ and $F^{\bd \to D}_0(m_{D}^2)$.

For pairs ``{\it b}'' and ``{\it c}'', the leading SU(3)-breaking
effect is the ratio of decay constants, $f_{D_s}/f_D$. This ratio has
been calculated on the lattice: $f_{D_s}/f_D = 1.22 \pm 0.04$
\cite{lattice}. Using this value, the error due to leading-order SU(3)
breaking is quite small. 

Although factorization has been shown to apply to $B \to D^{(*)} {\bar
D}^{(*)}$ decays \cite{luorosner}, it is also important to consider
SU(3) breaking in nonfactorizable contributions. Evidence of
factorization in $B \to D^{(*)} X$ can be understood as a consequence of large
$N_c$ QCD \cite{lukewise}: nonfactorizable corrections in $B \to
D^{(*)} {\bar D}^{(*)}$ decays then arise at $O(1/N_c^2)$. SU(3) breaking
in such effects is quite small: $\sim 25\%$ of $1/N_c^2$ is $\sim
3\%$. Therefore, the principal contribution to SU(3) breaking comes
from factorizable contributions [Eq.~(\ref{su3BDD})].

\subsection{$\bd \to \pi^+ \pi^-$}

The two potential partner processes to $\bd \to \pi^+ \pi^-$ are $\bs
\to K^0 {\bar K}^0$ and $B_u^+ \to K^0 \pi^+$. In order to estimate
the size of SU(3) breaking in these pairs of decays, we must calculate
matrix elements for nonleptonic $B$ decays. We use QCD factorization
\cite{BBNS} to do this. In this framework, all amplitudes are
calculated in the $m_b \to \infty$ limit. The corrections are then
$O(1/m_b)$. In this subsection, we present this in some detail; the
description in subsequent subsections is more cursory.

The starting point for the calculation of hadronic $B$ matrix elements
is the SM effective hamiltonian for $B$ decays \cite{BuraseffH}:
\bea
H_{eff}^q &=& {G_F \over \protect \sqrt{2}}
[V_{ub}V^*_{uq}(c_1 O_1^q + c_2 O_2^q) \nn\\
&& \qquad - \sum_{i=3}^{10}(V_{ub}V^*_{uq} c_i^u
+V_{cb}V^*_{cq} c_i^c +V_{tb}V^*_{tq} c_i^t) O_i^q] + h.c.,
\label{Heff}
\eea
where the superscript $u$, $c$, $t$ indicates the internal quark, and
$q$ can be either a $d$ or $s$ quark. The operators $O_i^q$ are
defined as
\bea
O_1^q &=& \bar q_\alpha \gamma_\mu L u_\beta\bar
u_\beta\gamma^\mu Lb_\alpha\;,\;\;\;\;\;\;O_f^q =\bar q
\gamma_\mu L u\bar
u\gamma^\mu L b\;,\nn\\
O_{3,5}^q &=&\bar q \gamma_\mu L b
\bar q' \gamma^\mu L(R) q'\;,\;\;\;\;\;\;\;O_{4,6}^q = \bar q_\alpha
\gamma_\mu Lb_\beta
\bar q'_\beta \gamma^\mu L(R) q'_\alpha\;,\\
O_{7,9}^q &=& {3\over 2}\bar q \gamma_\mu L b  e_{q'}\bar q'
\gamma^\mu R(L)q'\;,\;O_{8,10}^q = {3\over 2}\bar q_\alpha
\gamma_\mu L b_\beta
e_{q'}\bar q'_\beta \gamma^\mu R(L) q'_\alpha ~, \nn
\eea
where $R(L) = 1 \pm \gamma_5$, and $q'$ is summed over $u$, $d$, $s$.
$O_2$ and $O_1$ are the tree-level and QCD-corrected operators,
respectively. $O_{3-6}$ are the strong gluon-induced penguin
operators, and operators $O_{7-10}$ are due to $\gamma$ and $Z$
exchange (electroweak penguins), and ``box'' diagrams at loop level.
The values of the various Wilson coefficients (WC's) are given by
\cite{BuraseffH}
\bea
c_1 = -0.185 &,& c_2 = 1.082 ~, \nn\\
c^t_3 = 0.014 ~,~~ c^t_4 = -0.035 &,& c^t_5 = 0.010 ~,~~
c^t_6 =-0.041 ~, \nn\\
c^t_7 = -1.24\times 10^{-5} ~,~~ c_8^t = 3.77\times 10^{-4} &,&
c_9^t = -0.010 ~,~~ c_{10}^t = 2.06\times 10^{-3} ~.
\label{wc}
\eea

We will split the contribution to the decay amplitudes in QCD
factorization into a factorizable piece and a nonfactorizable piece.
This latter part includes the hard spectator interactions and
annihilation-type contributions which are formally suppressed by
$1/m_b$. We should point out that it is possible that there are
additional nonfactorizable contribution to nonleptonic decays, of
order $\Lambda_{QCD}/m_b$, which are missed in the QCD factorization
approach \cite{charming}. However the size and methods to calculate
such contributions remain a contentious issue and so we will not
include them in our estimates of the amplitudes. In any event, the
SU(3) breaking from these terms is suppressed by the heavy $b$-quark
mass, and will be small.
 
We write the general amplitudes $A_{s,d}$ for $\btos$ and $\btod$
penguin processes as
\beq
A_{s,d} = A_{s,d}^{fac}+ A_{s,d}^{nonfac} ~,
\eeq
where $A_{s,d}^{fac}$ is the factorization contribution and
$A_{s,d}^{nonfac}$ is the additional nonfactorizable contribution.
In relating $A_s$ and $A_d$, we write
\bea
A_s^{fac} &=& (1+z_{fac})A_d^{fac} ~, \nn\\
A_s^{nonfac} &=& (1+z_{nonfac})A_d^{nonfac} ~,
\eea
where $z_{fac}$ and $z_{nonfac}$ are the SU(3)-breaking corrections
which are typically $\sim 25\%$. The net SU(3) breaking in the
amplitude is then
\bea
\frac{A_s^{fac}+A_s^{nonfac}}{A_d^{fac}+A_d^{nonfac}}
& = &
1+z_{fac} + (z_{nonfac}-z_{fac})\frac{A_d^{nonfac}}
{A_d^{fac}+A_d^{nonfac}} ~.
\label{su3}
\eea
If $z_{nonfac} \sim z_{fac}$, then $z_{fac}$ clearly gives the
estimate of SU(3) breaking even in the presence of non-negligible
nonfactorizable effects. And if $A_d^{nonfac} \ll A_d^{fac}$, then
$z_{fac}$ again gives the estimate of SU(3) breaking. We will use
Eq.~(\ref{su3}) to estimate SU(3) breaking from nonfactorizable
corrections.

In order to identify the sources of SU(3) breaking, we note that the
factorizable contributions to $\bd \to \pi^+ \pi^-$ and $\bs\to K^0
\kbar$ can be written as
\bea
A(\bd \to \pi^+ \pi^-) & = & f_{\pi} \, F^{\bd \to \pi} \, 
\int T(x) \phi_{\pi}(x) dx ~ + 0(1/m_b) ~, \nn\\
A(\bs \to K^0 {\bar{K}}^0) & = & f_K \, F^{\bs \to \bar{K}} \, 
\int T(x) \phi_K(x) dx + 0(1/m_b) ~.
\label{fact1}
\eea
In the above, $F^{\bd \to \pi} $ and $F^{\bs \to \bar{K}}$ are form
factors, while the integrals represent the hadronization of quarks
into a $\pi$ or a $K^0$. The quantities $\phi_{\pi, K}$ are the pion
and kaon light cone distributions (LCD's), which can be expanded in
terms of Gegenbauer polynomials $C^{3/2}_n$ as follows \cite{Ballv}:
\beq
\phi_M(x,\mu) = f_K \, 6x(1-x) \left( 1 + \sum_{n=1}^\infty
\alpha_{n}^M(\mu) C_{2n}^{3/2}(2x-1) \right) ~,
\eeq
where the Gegenbauer moments $a_{2n}^M$ are multiplicatively
renormalized, change slowly with $\mu$, and vanish as $\mu \to
\infty$. The pion LCD is symmetric under $x \to 1-x$ because of
isospin symmetry. For the kaon the antisymmetric part of the LCD
arises from SU(3) breaking. That is, it is the presence of the
antisymmetric piece at scale $\mu \sim m_b$, proportional to odd
powers of $(2x - 1)$, which will generate SU(3) corrections from the
final-state kaons.

In what follows, we denote $\bd \to \pi^+ \pi^-$ and $\bs \to K^0
{\bar K}^0$ as pair ``{\it a}'', and $\bd \to \pi^+ \pi^-$ and $B_u^+
\to K^0 \pi^+$ as pair ``{\it b}''. Following the notation in
Ref.~\cite{BBNS} the amplitudes for all three decays can be explicitly
written as:
\bea
A( \bd \to \pi^+ \pi^-) & = & \Vud A_{ut}^{\pi \pi} +\Vcd A_{ct}^{\pi
\pi} ~; \nn\\
A_{ut}^{\pi \pi} & = & - \left[ a_2 + a_{4 \pi}^{u} +a_{10
\pi}^u+r_{\chi}^{\pi} (a_{6\pi}^u+a_{8 \pi}^u) \right ]A_{\pi \pi} ~,
\nn\\
A_{ct}^{\pi \pi} & = & - \left[ a_{4 \pi}^c +a_{10
\pi}^c+r_{\chi}^{\pi}(a_{6\pi}^c+ a_{8 \pi}^c) \right ]A_{\pi \pi} ~,
\nn\\
A_{\pi \pi} & = & \frac{iG_F}{\sqrt{2}}(m_B^2-m_{\pi}^2) F_0^{\bd \to
\pi}(m_{\pi}^2) f_{\pi} ~.
\label{Bpipi}
\eea
\bea
A( \bs \to K^0 {\bar K}^0 ) & = & \Vus A_{ut}^{KK} +\Vcs A_{ct}^{KK}
~; \nn\\
A_{ut}^{KK} & = & \left[ a_{4 K}^u -\frac{1}{2}a_{10
K}^u+r_{\chi}^{K} (a_{6 K}^u-\frac{1}{2}a_{8 K}^u) \right ]A_{K K} ~,
\nn\\
A_{ct}^{KK} & = & \left[ a_{4 K}^c - \frac{1}{2}a_{10
K}^c+r_{\chi}^{K}(a_{6 K}^c-\frac{1}{2}a_{8 K}^c) \right ] A_{KK} ~,
\nn\\
A_{K K} & = & \frac{iG_F}{\sqrt{2}}(m_{B_s}^2-m_{K}^2) F_0^{\bs \to
\bar{K}}(m_{K}^2) f_{K} ~.
\label{BKK}
\eea
\bea
A( B^+ \to K^0 \pi^+) & = & \Vus A_{ut}^{ K \pi} +\Vcs A_{ct}^{K \pi}
~; \nn\\
A_{ut}^{ K \pi} & = & \left[ a_{4 K}^u -\frac{1}{2}a_{10
K}^u+r_{\chi}^{K} (a_{6 K}^u-\frac{1}{2}a_{8 K}^u) \right ]A_{K \pi}
~, \nn\\
A_{ct}^{ K \pi} & = & \left[ a_{4 K}^c -\frac{1}{2}a_{10
K}^c+r_{\chi}^{K} (a_{6 K}^c-\frac{1}{2}a_{8 K}^c) \right ] A_{K \pi}
~, \nn\\
A_{K \pi} & = & \frac{iG_F}{\sqrt{2}}(m_{B}^2-m_{\pi}^2) F_0^{\bd \to
\pi}(m_{K}^2) f_{K} ~.
\label{BKpi}
\eea

In the above, the $a_i$ are combinations of WC's and are given by $a_i
= c_i + c_{i+1}/N_c$ ($i$ odd) or $a_i = c_i + c_{i-1}/N_c$ ($i$
even). The chiral enhancement terms are given by
\beq
r_{\chi}^{\pi}=\frac{2m_{\pi}^2}{m_b(m_u+m_d)} ~~,~~~~
r_{\chi}^{K}=\frac{2m_{K}^2}{m_b(m_u+m_s)} ~.
\label{chiralenhance}
\eeq
The form factors are defined in Eq.~(\ref{formfactor}). Note that both
$\bd \to \pi^+ \pi^-$ and $B^+ \to K^0 \pi^-$ involve the same form
factor $F_0^{\bd \to \pi}$. This follows from isospin symmetry.

Our methods assume the SU(3) relations
\beq
A_{ct}^{ \pi \pi} = A_{ct}^{K K} = A_{ct}^{ K \pi} ~.
\label{input}
\eeq
{}From Eqs.~(\ref{Bpipi}), (\ref{BKK}) and (\ref{BKpi}), we can thus
identify three possible sources of SU(3) breaking: (i) $\pi$ vs.\
$K$ hadronization. This is represented by differences between the
$a_{iK}$ and $a_{i \pi}$, which is related to differences in the $\pi$
and $K$ LCD's; (ii) the difference in form factors; (iii) differences
in the chiral enhancement factors $ r_{\chi}^{\pi}$ and $r_{\chi}^K$.

We are now able to compute the SU(3) breaking as
\bea
1+z^a_{fac} &= & \frac {A_{ct}^{ K K}}{A_{ct}^{\pi \pi}}= \frac{
\left[ a_{4 K}^c -\frac{1}{2}a_{10 K}^c+r_{\chi}^{K} (a_{6 K}^c
-\frac{1}{2}a_{8 K}^c) \right ] A_{KK} } { \left[ a_{4 \pi}^c +a_{10
\pi}^c+r_{\chi}^{\pi} (a_{6 \pi}^c+a_{8 \pi}^c) \right ]A_{\pi \pi}}
~, \nn\\
1+z^b_{fac} &= & \frac{A_{ct}^{ K \pi}}{A_{ct}^{\pi \pi}} = \frac{
\left[ a_{4 K}^c -\frac{1}{2}a_{10 K}^c+r_{\chi}^{K}(a_{6 K}^c -
\frac{1}{2}a_{8 K}^c) \right ]A_{K \pi} } { \left[ a_{4 \pi}^c +a_{10
\pi}^c+r_{\chi}^{\pi}(a_{6 \pi}^c+ a_{8 \pi}^c) \right ] A_{\pi \pi} }
~.
\label{su3ratiosfull}
\eea

{}From the values of the WC's in Eq.~(\ref{wc}), we note that the
contributions from the color-suppressed electroweak penguins
represented by $a_{8,10}$ are tiny. Neglecting these, we can simplify
$z^{a,b}_{fac}$ as
\bea
1+z^a_{fac} &= & \frac {A_{ct}^{ K K}}{A_{ct}^{\pi \pi}}= \frac{
\left[ a_{4 K}^c +r_{\chi}^{K} a_{6 K}^c \right ] A_{KK} }
{ \left[ a_{4 \pi}^c ++r_{\chi}^{\pi}a_{6 \pi}^c \right ]A_{\pi \pi} }
= \tilde{a}_{ K \pi} \frac{A_{KK}}{A_{\pi \pi}} ~, \nn\\
1+z^b_{fac} &= & \frac{A_{ct}^{ K \pi}}{A_{ct}^{\pi \pi}} = \frac{
\left[ a_{4 K}^c +r_{\chi}^{K}a_{6 K}^c \right ]A_{K \pi} }
{ \left[ a_{4 \pi}^c +r_{\chi}^{\pi}a_{6 \pi}^c \right ] A_{\pi \pi} }
= \tilde{a}_{ K \pi} \frac{A_{K \pi}}{A_{\pi \pi}} ~.
\label{su3simple}
\eea
In the above, $\tilde{a}_{K \pi}$ represents the SU(3) breaking in the
$a_i$'s due to the different LCD's of the kaon and the pion, as well
as in $r^K_{\chi}/r^{\pi}_{\chi}$. If the pion and the kaon LCD's take
their asymptotic form, then, taking the estimate of
$r^K_{\chi}/r^{\pi}_{\chi}=0.99 \pm 0.06$ from
Ref.~\cite{BsKKSU3break}, we see that to a very good approximation
$\tilde{a}_{K \pi} \approx 1$. The recent measurement of the pion LCD
at $\mu^2\sim 10~{\rm GeV}^2$ \cite{pionLCD} shows that the pion LCD
is extremely close to its asymptotic form, $\phi_\pi(x) \sim
x(1-x)$. (Note that isospin symmetry requires only that the pion LCD
be symmetric, not asymptotic.) This suggests that, at the scale $\mu
\sim m_b$, the LCD's of the light mesons $K$ and $K^*$ are probably
also very close to their asymptotic form, i.e.\ symmetric under the
interchange $x \leftrightarrow 1-x$. Allowing for SU(3) breaking in
the kaon LCD we will take the first Gegenbauer moment to be equal to
$\alpha_{1}^K=0.2 \pm 0.02$ \cite{BsKKSU3break}. The maximum SU(3)
breaking then corresponds to $\tilde{a}_{K \pi}=0.973$, and is only
3\%.

One can also consider the vector-vector decays $\bd \to \rho^+
\rho^-$, $\bs \to K^{*0} {\bar K}^{*0}$ and $B_u^+ \to K^{*0}
\rho^+$. In this case the corresponding value for $\tilde{a}_{K^*
\rho}$ is $\tilde{a}_{K^* \rho}=0.963$, which is only a 4\% SU(3)
breaking. Note that the SU(3) breaking in the $K$ and $K^*$ LCD's from
$\alpha_1^{K(K^*)}$ is very similar: model calculations find the
antisymmetric piece to be equal at the level of $\sim 10\%$
\cite{Ballv}.

It is clear from the above discussion that the SU(3) breaking in the
meson LCD's is small. Henceforth we will neglect these corrections and
assume the asymptotic form for the LCD's of the various mesons. We can
therefore write
\bea
1+z^a_{fac} & \approx & \frac{A_{KK}} {A_{\pi \pi}}=
\frac{(m_{B_s}^2-m_{K}^2) F_0^{\bs \to \bar{K}}(m_{K}^2)f_K }
{(m_{B}^2-m_{\pi}^2) F_0^{\bd \to \pi}(m_{\pi}^2)f_{\pi} } \approx
\frac{ F_0^{\bs \to \bar{K}}(m_{K}^2)f_K } { F_0^{\bd \to
\pi}(m_{\pi}^2)f_{\pi} } ~,
\label{su3finalfaca} \\
1+z^b_{fac} & \approx & \frac{A_{K \pi}} {A_{\pi \pi}}= \frac{
F_0^{\bd \to \pi}(m_{K}^2)f_K } { F_0^{\bd \to \pi}(m_{\pi}^2)f_{\pi}}
\approx \frac{f_K}{f_{\pi}} ~.
\label{su3finalfacb}
\eea
In the second line, we have assumed that the $\bd \to \pi$ form factor
is similar at $q^2=m_K^2$ and $q^2=m_{\pi}^2$. This is reasonable: in
$\bd \to \pi$ transitions, where the energy transferred is large, the
difference of form factors at $q^2=m_{K}^2$ and $m_{\pi}^2$ is
suppressed by some power of $m_{K,\pi}^2/m_B^2$ and is therefore
negligible. 

The factorizable contributions to SU(3) breaking in the decay pairs
$\bd \to \pi^+ \pi^-$ and $\bs \to K^0 {\bar K}^0$, and $\bd \to \pi^+
\pi^-$ and $B_u^+ \to K^0 \pi^+$, are given by the above expressions.
Within QCD factorization, there are (unknown) $1/m_b$ corrections to
these results. Their size is typically $(m_s/\Lambda_{QCD}) \times
(\Lambda_{QCD}/m_b) = m_s/m_b \sim 5\%$.

We now turn to the nonfactorizable contributions. An example of such
an effect is the correction due to hard gluon exchange between the
spectator quark and the energetic quarks of the emitted meson. We can
parametrize these corrections as
\bea
A_{non fac}^{M_1 M_2} & = & f_B f_{M_1}f_{M_2} \frac{m_B}{\lambda_B} 
B^H_{M_1 M_2} ~,
\label{hard}
\eea
where $m_{B_q}/\lambda_{B_q} = \int \phi_{B_q}(z)/z$, in which
$\phi_{B_q}(z)$ is the $B^0_q$ LCD, $q=d,s$. The quantity $B^H_{M_1
M_2}$ depends on the final state and cannot be calculated as it
suffers from endpoint divergences that will eventually be smoothed out
by unknown soft physics. In the approach of Ref.~\cite{BBNS}, the
divergent piece is regulated by an unknown parameter which is then
assumed to be within a certain range. Here we adopt this same method.

We can now estimate the size of nonfactorizable SU(3) breaking in the
pairs of processes ``{\it a}'' and ``{\it b}'':
\bea
1+z^a_{non fac} & = & \frac {A_{ct}^{ K K}}{A_{ct}^{\pi \pi}} =
\frac{f_{B_s}}{f_{B_d}} \frac{\lambda_{B_d}}{\lambda_{B_s}}
\frac{f_K^2}{f_{\pi}^2} \frac{m_{B_s}}{m_B} \frac{B^H_{KK}}{B^H_{\pi
\pi}} ~, \nn\\
1+z^b_{non fac} & = & \frac {A_{ct}^{ K \pi}}{A_{ct}^{\pi \pi}} =
\frac{f_K}{f_{\pi}} \frac{B^H_{K\pi}}{B^H_{\pi \pi}} ~.
\label{su3hard}
\eea
There are two sources of SU(3) breaking. The first comes from the
ratios of the $B^H_{M_1 M_2}$. We assume that both $B^H_{KK}/B^H_{\pi
\pi}$ and $B^H_{K \pi}/B^H_{\pi \pi}$ are $\sim 1 \pm 0.25$. (The
error of 25\% (typical of SU(3) breaking) is somewhat larger than the
estimate using the approach in Ref. \cite{BBNS}.) The second comes
from the initial state. This SU(3) breaking, $z_i$, is given by
\beq
z_i = \frac{f_{B_s}}{f_{B_d}} \frac{\lambda_{B_d}}{\lambda_{B_s}} ~.
\label{su3initial}
\eeq
In a simple model one can write $f_{B_q} = \mu_q^{3/2}/m_{B_q}^{1/2}$
and $\lambda_{B_q} \sim \mu_q$, where $\mu_q$ is the reduced mass,
which is different for the $\bs$ and the $\bd$ mesons. In the
heavy-quark limit we have $f_{B_s}/f_{B_d} =\mu_s^{3/2}/\mu_d^{3/2}$
and $(f_{B_s}/\lambda_{B_s}) / (f_{B_d}/\lambda_{B_d})
=\mu_s^{1/2}/\mu_d^{1/2}$. Taking $f_{B_s}/f_{B_d} = 1.15$, we find
that the initial-state SU(3)-breaking correction $z_i-1$ is $\sim
5\%$.

Using Eqs.~(\ref{su3}), (\ref{su3finalfaca}) and (\ref{su3finalfacb}),
we can then write the total SU(3)-breaking correction as
\bea
z^a_{total} & \!\!=\!\! & \frac {A_{ct}^{ K K}}{A_{ct}^{\pi \pi}}-1 \approx
z^a_{fac}-(1+z^a_{fac})\left[1- z_i\frac{f_K}{f_{\pi}} \frac{ F_0^{\bd
\to \pi}(m_{\pi}^2) } { F_0^{\bs \to \bar{K}}(m_{K}^2) }
\frac{B^H_{KK}}{B^H_{\pi \pi}} \right] \frac{A_{non fac}^{\pi \pi}}
{A_{ fac}^{\pi \pi}+{A_{non fac}^{\pi \pi}}} \nn\\
z^b_{total} & \!\!=\!\! & \frac {A_{ct}^{ K \pi}}{A_{ct}^{\pi \pi}}-1 \approx
z^b_{fac}-(1+z^b_{fac})\left[1- \frac{B^H_{K\pi}}{B^H_{\pi \pi}}
\right] \frac{A_{non fac}^{\pi \pi}} {A_{ fac}^{\pi \pi}+{A_{non
fac}^{\pi \pi}}} ~.
\label{su3total}
\eea
The calculation of Ref.~\cite{BBNS} indicates that the hard spectator
correction is small, typically about 10\% or less in the decays we are
considering. Combined with the fact that we expect $(f_K/f_{\pi})
(F_0^{\bd \to \pi}(m_{\pi}^2)/F_0^{\bs \to \bar{K}}(m_{K}^2)) \sim 1$
in the second term in $z^a_{total}$ above, we find that the SU(3)
corrections from the hard spectator corrections are $\sim 3\%$, which
is quite small.

We therefore conclude that, to the extent that exchange- and
annihilation-type topologies are unimportant, the principal
SU(3)-breaking effect in $\bd \to \pi^+ \pi^-$ and $\bs \to K^0 {\bar
K}^0$ or $\bd \to \pi^+ \pi^-$ and $B_u^+ \to K^0 \pi^+$ comes from
the factorizable contributions to the decays
[Eqs.~(\ref{su3finalfaca}) and (\ref{su3finalfacb})]. 

\subsection{$\bd \to K^0 {\bar K}^0$}

There are two candidate partner processes for $\bd \to K^0 {\bar
K}^0$: $\bs \to K^0 {\bar K}^0$ and $B_u^+ \to K^0 \pi^+$. Below we
analyze separately the SU(3) breaking in each of these pairs of
decays.

We consider first $\bd \to K^0 {\bar K}^0$ and $\bs \to K^0 {\bar
K}^0$ \cite{BKKbar}. Following the analysis of the previous section we
estimate the SU(3) breaking in this pair of decays to be
\bea
1+z_{fac} & \approx & \frac{A_{KK}^s} {A_{K K}^d}=
\frac{ F_0^{\bs \to \bar{K}}(m_{K}^2) } { F_0^{\bd \to K}(m_{K}^2) } ~.
\label{su3KK}
\eea
Since the final states are the same for both decays, there is no SU(3)
breaking due to decay constants. 

In the case of second pair of decay processes, $\bd \to K^0 {\bar
K}^0$ and $B_u^+ \to K^0 \pi^+$, the analysis of the previous
subsection gives
\beq
1+z_{fac} \approx \frac{A_{K \pi}} {A_{K K}}=
\frac{ F_0^{\bd \to \pi}(m_{K}^2) } { F_0^{\bd \to K}(m_{K}^2) } ~.
\label{su3piKKK}
\eeq

\subsection{$\bd\to\rho^0\rho^0$}

The partner process to $\bd\to\rho^0\rho^0$ is $\bd\to K^{*0} \rho^0$.
Since in this case we are dealing with vector-vector final states, we
have to consider specific helicity states. In the linear polarization
basis there are three independent polarization amplitudes. They are
$A_0$ (longitudinal amplitude) and $A_{\|, \perp}$ (two transverse
amplitudes). Ignoring small differences in the WC's for the two
processes, the SU(3)-breaking term for a given polarization state
$\lambda= 0, \perp, \|$ is given by
\beq
1+z_{fac}^\lambda = {\left[ a_4^c - {3\over 2} a_9^c - {1\over
2}a_{10}^c \right] A_{\rho\rho}^\lambda \over \left[ a_4^c - {1\over
2}a_{10}^c \right] A_{K^* \rho}^\lambda + \left[ - {3\over 2} a_9^c
\right] A_{\rho K^*}^\lambda } ~.
\label{su3rhorho}
\eeq
Within QCD factorization, one must now express the polarization
amplitudes $A_{\rho\rho}^\lambda$, $A_{K^* \rho}^\lambda$ and $A_{\rho
K^*}^\lambda$ in terms of masses, decay constants and form factors. We
do this below.

The polarization amplitudes $A_{V_1 V_2}^\lambda$ for the process $B
\to V_1 V_2$ are given as \cite{dattalondonTP}:
\bea
A_{V_1V_2}^\| & = & \sqrt{2} \, a_{V_1V_2} \nn\\ 
A_{V_1V_2}^0 & = & -a_{V_1V_2} x - {m_1 m_2 \over m_{\sss B}^2} \,
b_{V_1V_2} (x^2 - 1) ~, \nn\\
A_{V_1 V_2}^\perp & = & 2\sqrt{2} \, {m_1 m_2 \over m_{\sss B}^2}
c_{V_1V_2} \sqrt{x^2 - 1} ~,
\label{Adefs}
\eea
with $x = k_1 \cdot k_2 / (m_1 m_2)$, where $k_{1,2}$ and $m_{1,2}$
are the momenta and the masses of the vector mesons $V_{1,2}$. The
parameters $a_{V_1V_2}$, $b_{V_1V_2}$ and $c_{V_1V_2}$ can be written
\cite{dattalondonTP}
\bea
a_{V_1V_2} &=& - m_1g_{V_1}(m_B+m_2)A_1^{(2)}(m_1^2) ~, \nn\\
b_{V1V_2} &=& 2m_1g_{V_1}{m_B \over (m_B+m_2)}m_B A_2^{(2)}(m_1^2) ~,
\nn\\
c_{V_1V_2} &=&-m_1g_{V_1}{m_B \over (m_B+m_2)}m_B V^{(2)}(m_1^2) ~.
\label{abc}
\eea
These quantities depend on the form factors $V^{(2)}$ and
$A_{1,2}^{(2)}$, and on the decay constant of $V_1$, $g_{V_1}$. These
are defined as follows. 

A $B\to V_i$ transition is described by the four form factors
$V^{(i)}$, $A_{0,1,2}^{(i)}$ \cite{BSW}:
\bea
\bra{V_i(k_i, \lambda)} {\bar q}' \gamma_\mu b \ket{B(p)}& =& i {{2
V^{(i)}(q^2)} \over (m_B+m_i)} \epsilon_{\mu \nu\rho\sigma} p^\nu
k_i^\rho \varepsilon_{\lambda}^{*\sigma} ~,\nn\\
\bra{V_i(k_i, \lambda)} {\bar q}' \gamma_\mu\gamma_5 b \ket{B(p)}& =&
(m_B+m_i)A_1^{(i)}(q^2) \left[\varepsilon_{\mu, \lambda}^{*}-
\frac{\varepsilon_{\lambda}^{*}.q}{q^2}q_{\mu}\right]\nn\\
&& ~-A_2^{(i)}(q^2)\frac{\varepsilon_{\lambda}^{*}.q}{m_B+m}
\left[(p_{\mu}+k_{\mu}^\lambda)-
\frac{m_B^2-m_i^2}{q^2}q_{\mu}\right]\nn\\
&& ~+2 m_i\frac{\varepsilon_{\lambda}^{*}.q}{q^2}q_{\mu}A_0^{(i)}(q^2)
~,
\label{ffactor}
\eea   
where $q=p-k_i$. Although the values of these form factors are
model-dependent, the number of independent form factors can be reduced
if the dominant contribution to the form factors comes from soft gluon
interactions between the quarks inside the mesons. In this case, in
the limit $m_b \to \infty$ and $E_V \to \infty$, one has the following
relations between the vector form factors \cite{Charles}:
\bea 
A_1(q^2 ) &=& \frac{2E_V}{ m_B +m_V} \zeta_{\perp}(m_B, E_V) ~, \nn\\
A_2(q^2 ) & = & ( 1+ \frac{m_V}{m_B}) \zeta_{\perp}(m_B, E_V)
\left[1-\frac{m_V}{E_V}\frac{\zeta_{\|}(m_B, E_V)}{\zeta_{\perp}(m_B,
E_V)}\right] ~, \nn\\
V(q^ 2) &=&( 1+ \frac{m_V}{m_B}) \zeta_{\perp}(m_B, E_ V) ~.
\label{leet}
\eea 
{}From this we see that the form factors $A_{1,2}$ and $V$ are
expressible in terms of two form factors $\zeta_{\perp,\|}$. Moreover,
we see that the difference between $\zeta_{\|}$ and $\zeta_{\perp}$ is
suppressed by $m_V/E_V$. Coupled with the fact that model calculations
indicate that $\zeta_{\|} \sim \zeta_{\perp}$, we can assume that
$\zeta_{\|} = \zeta_{\perp}$ for the purpose of calculating SU(3)
breaking.

The decay constant of a vector mesons is defined by $m_V g_V
\epsilon_{\mu}^* \sim \bra{V(p, \epsilon)} J_{\mu} \ket{0}$. However,
the size of the polarization vector $\epsilon_{\mu}^*$ differs for
different polarizations: for transverse polarization $\epsilon \sim
0(1)$, while for longitudinal polarization $\epsilon \sim E_V / m_V$.
This must be taken into account when evaluating the $A_{V_1 V_2}$.

Putting all of the above information together, we can obtain
expressions for $A_{\rho\rho}^\lambda$, $A_{K^* \rho}^\lambda$ and
$A_{\rho K^*}^\lambda$. Note that $A_{K^* \rho}^\lambda \sim g_{K^*}
F^{\bd \to \rho}$ and $A_ {\rho K^*}^\lambda \sim g_\rho F^{\bd \to
K^*}$, where $F^{\bd \to \rho}$ and $F^{\bd \to K^*}$ are form
factors. Since $F^{\bd \to K^*} / F^{\bd \to \rho} \sim g_{K^*} /
g_{\rho}$, we have $A_{K^* \rho}^\lambda \simeq A_{\rho
K^*}^\lambda$. Combined with the fact that the EWP WC $a_9^c$ is
smaller than the QCD penguin WC $a_4^c$, we can simplify
Eq.~\ref{su3rhorho} as
\beq
1+z_{fac}^\lambda = {\left[ a_4^c - {3\over 2} a_9^c - {1\over
2}a_{10}^c \right] A_{\rho \rho}^\lambda \over \left[ a_4^c - {3\over
2} a_9^c- {1\over 2}a_{10}^c \right] A_{K^*\rho}^\lambda + {3\over 2}
a_9^c \left[A_{\rho K^*}^\lambda- A_{\rho K^*}^\lambda\right] } 
\approx { A_{\rho \rho}^\lambda \over A_{K^* \rho}^\lambda} ~. 
\label{su3rhorhoapprox}
\eeq

Both $A_{\rho \rho}^\lambda$ and $A_{K^* \rho}^\lambda$ involve the
same $B \to\rho$ form factors, so that here SU(3) breaking depends
only on the ratios of decay constants and masses of the $K^*$ and
$\rho$. This can be seen easily as follows: neglecting terms of
$O(m_{\rho,K^*}^2/m_B^2)$, the expressions for the polarization
amplitudes simplify \cite{dattalondonTP}:
\bea
A_{V_1V_2}^\| & = & \sqrt{2} \, a_{V_1V_2} \nn\\ 
A_{V_1V_2}^0 & = & -(2a_{V_1V_2} + b_{V_1V_2}) \frac{m_B^2}{4m_1m_2}
~, \nn\\
A_{V_1 V_2}^\perp & = & \sqrt{2} \, c_{V_1V_2} ~.
\label{smallmass}
\eea
The SU(3)-breaking effects for the different polarization states are
then given as
\beq
1+z_{fac}^0 = \frac{g_{\rho}}{g_{K^*}} ~~;~~~~
1+z_{fac}^{\perp, \|} = \frac{m_{\rho}g_{\rho}}{m_{K^*}g_{K^*}} ~.
\label{su3rhorhopol}
\eeq
 
\subsection{$\bs \to {\bar K}^{*0} \rho^0$}

Here there is a single candidate partner process: $\bd \to
K^{*0}\rho^0$. The SU(3) breaking is given by
\beq
1+z_{fac}^\lambda = {\left[ a_4^c - {3\over 2} a_9^c - {1\over
2}a_{10}^c \right] A_{\rho K^*}^{s,\lambda} \over \left[ a_4^c - {1\over
2}a_{10}^c \right] A_{K^* \rho}^\lambda + \left[ - {3\over 2} a_9^c
\right] A_{\rho K^*}^\lambda } ~,
\label{su3Kstarrho}
\eeq
where $A_{K^* \rho}^\lambda$ and $A_{\rho K^*}^\lambda$ are given in
Eq.~(\ref{Adefs}), and $A_{\rho K^*}^{s,\lambda}$ is obtained from
$A_{\rho K^*}^{\lambda}$ by replacing $\bd$ by $\bs$. As before we can
approximate the SU(3) breaking as
\beq
1+z_{fac}^\lambda \approx {A_{\rho K^*}^{s,\lambda} \over
A_{K^*\rho}^\lambda} ~.
\eeq
Using the expressions given in the previous subsection, this yields
\bea
1+z_{fac}^0 & = & {m_{\bs}^2 \over m_{\bd}^2} {m_\rho \, g_\rho \over
m_{K^*} \, g_{K^*}} {(m_{\bd} + m_\rho) \over (m_{\bs} + m_{K^*})} \times \nn\\
& & \quad {-(m_{\bs} + m_{K^*})^2 \, A_1^{\bs\to K^*}(m_\rho^2) + m_{\bs}^2
\, A_2^{\bs\to K^*}(m_\rho^2) \over -(m_{\bd} + m_\rho)^2 \,
A_1^{\bd\to \rho}(m_{K^*}^2) + m_{\bd}^2 \, A_2^{\bd\to
\rho}(m_{K^*}^2)} ~, \nn\\
1+z_{fac}^\| & = & { m_\rho \, g_\rho \, (m_{\bs} + m_{K^*}) \,
A_1^{\bs\to K^*}(m_\rho^2) \over m_{K^*} \, g_{K^*} \, (m_{\bd} +
m_\rho) \, A_1^{\bd\to \rho}(m_{K^*}^2)} ~, \nn\\
1+z_{fac}^\perp & = & { m_\rho \, g_\rho \, m_{\bs}^2 \, (m_{\bd} +
m_\rho) \, V^{\bs\to K^*}(m_\rho^2) \over m_{K^*} \, g_{K^*} \,
(m_{\bs} + m_{K^*}) \, V^{\bd\to \rho}(m_{K^*}^2)} ~.
\label{su3Kstarrhoapprox}
\eea
We see that here the SU(3)-breaking term depends on form factors, as
well as decay constants and vector-meson masses.

\subsection{$\bs \to \phi {\bar K}^{*0}$}

In this case there are three candidate partner processes: $\bd \to
\phi K^{*0}$, $B_u^+ \to \phi K^{*+}$ and $\bs \to \phi\phi$. We refer
to the three pairs of decays, in this order, as pair ``{\it a}'',
``{\it b}'' and ``{\it c}''. The SU(3)-breaking term for both pairs
``{\it a}'', and ``{\it b}'' is given by
\bea
1+z_{fac}^\lambda = {\left[a_3^c + a_4^c +a_5^c - {1 \over 2} a_7^c -
{1 \over 2} a_9^c - {1\over 2}a_{10}^c \right] A_{\phi K^*}^\lambda
\over \left[a_4^c - {1\over 2}a_{10}^c \right] A_{{\bar
K}^*\phi}^{s,\lambda} + \left[a_3^c+a_5^c - {1 \over 2} a_7^c- {1\over
2} a_9^c \right] A_{\phi {\bar K}^*}^{s,\lambda}} ~.
\label{su3Kstarphi}
\eea
The SU(3) breaking for ``{\it c}'' has the same form as the above, but
with the $A_{\phi K^*}^\lambda$ in the numerator being replaced by
$A_{\phi\phi}^{s,\lambda}$. Previous arguments tell us that
$A_{K^*\phi}^{s,\lambda}$ and $A_{\phi K^*}^{s,\lambda}$ are of
similar size. Thus, using the fact that the WC's $a_{3,5}^c$ and
$a_{7,9}^c$ are much smaller than the QCD penguin WC $a_4^c$, we can
rewrite Eq.~\ref{su3Kstarphi} as
\beq
1+z^{a,\lambda}_{fac} \approx 1+z^{b,\lambda}_{fac} \approx
\frac{A_{\phi K^*}^\lambda}{A_{{\bar K}^* \phi}^{s,\lambda}} ~~;~~~~
1+z^{c,\lambda}_{fac} \approx
\frac{A_{\phi\phi}^{s,\lambda}}{A_{{\bar K}^* \phi}^{s,\lambda}} ~.
\label{su3Kstarphiabc}
\eeq
For pairs ``{\it a}'' and ``{\it b}'', the SU(3)-breaking term
includes both decay constants and form factors, as in
Eq.~(\ref{su3Kstarrhoapprox}). However, for pair ``{\it c}'', all
dependence on form factors vanishes, and we have
\beq
1+z_{fac}^{c,0} = \frac{g_{\phi}}{g_{K^*}} ~~;~~~~
1+z_{fac}^{c,\perp, \|} = \frac{m_{\phi}g_{\phi}}{m_{K^*}g_{K^*}} ~.
\eeq

\subsection{Annilation and Exchange Contributions}

The amplitudes for most pairs of decays discussed previously are equal
in the SU(3) limit only if annihilation- and exchange-type amplitudes
are neglected. Thus, for these decays, it is unnecessary to consider
SU(3) breaking in such contributions. Note that, as mentioned earlier,
this assumption can be tested experimentally. If it turns out that
such amplitudes are large, perhaps because of chiral enhancements,
then one can use instead vector-vector final states for which
annihilation and exchange contributions are expected to be small.

There are, however, three pairs of decays whose amplitudes are equal
in the SU(3) limit: $\bd \to D^+ D^-$ and $\bs \to D_s^+ D_s^-$, $\bd
\to K^0 {\bar K}^0$ and $\bs \to K^0 {\bar K}^0$, and $\bs \to \phi
{\bar K}^{*0}$ and $\bd\to\phi K^{*0}$. For these decays, one need not
neglect annihilation and exchange amplitudes. In this case, one can
consider the size of SU(3)-breaking effects in such contributions.
Below, we present an estimate of this SU(3) breaking within QCD
factorization for $\bd \to K^0 {\bar K}^0$ and $\bs \to K^0 {\bar
K}^0$.

The annihilation terms for these decays can be parametrized as
\bea
A_{ann}^{s,\bar{K} K} & = & f_{B_s} f_{K}f_{K} X^s_{KK} ~, \nn\\
A_{ann}^{d,\bar{K} K} & = & f_{B_d} f_{K}f_{K} X^d_{KK} ~.
\label{annKK}
\eea
Like the hard spectator corrections, the quantities $X^{s,d}_{KK}$
depend on the kaon LCD and are divergent because of missing soft
contributions. Within the QCD factorization approach
\bea
X^{s,d}_{KK} \sim r_{\chi}^{K}(2X^2_{s,d}-X_{s,d}) ~,
\label{annX}
\eea
where the divergent quantities $X_{s,d}$ are regulated as
\bea
X_{s,d} = (1+\rho_{s,d}e^{i\phi_{s,d}}) \, \ln\frac{m_B}{\Lambda_{QCD}} ~.
\eea
$\rho_s$ and $\rho_d$ are related by SU(3), as are $\phi_s$ and
$\phi_d$. However, these quantities are not known; they are taken to
satisfy $|\rho_{s,d}| <1$ \cite{BBNS}. The SU(3) breaking in
Eq.~(\ref{annKK}) can then be estimated by assuming that $\rho_{d,s}$
and $\phi_{d,s}$ differ by 25\%.

Thus, including annihilation terms, the size of the SU(3) corrections
can be estimated to be
\beq
z_{total} = \frac {A_{s,ct}^{ K K}}{A_{d,ct}^{K K}}-1 \approx
z_{fac}-(1+z_{fac})\left[1- \frac{f_{B_s}}{f_{B}} \frac{ F_0^{\bd
\to {K}^0}(m_{K}^2) } { F_0^{\bs \to \bar{K}}(m_{K}^2) }
\frac{X^s_{KK}}{X^d_{K K}} \right]R_A ~,
\label{annKKfinal}
\eeq
where
\beq
R_A = \frac{A_{ann}^{d, K K}} {A_{ fac}^{d,K K}+{A_{ann}^{d,K K}}} ~.
\eeq
The total amount of SU(3) breaking clearly depends on the size of the
annihilation contributions. If the annihilation terms are large ---
Ref.~\cite{PQCD} estimates them to be $\sim 40\%$ --- then the SU(3)
breaking due to such terms could be 10-15\%.

\section{Discussion}

Our findings are summarized in Table~\ref{summarytable}.  There are
six neutral $B^0 \to M_1 M_2$ decays involving a $\btod$ penguin
amplitude in which both $B^0$ and $\Bbar$ mesons can decay to $M_1
M_2$. For each of these decays, we list the potential partner
processes $B' \to M'_1 M'_2$ which receive a significant $\btos$
penguin contribution and which are dominated by a single amplitude.
There are a total of twelve decay pairs to which our methods can be
applied.

\TABLE{
\hfil
\vbox{\offinterlineskip
\halign{&\vrule#&
 \strut\quad#\hfil\quad\cr
\noalign{\hrule}
height2pt&\omit&&\omit&&\omit&&\omit&\cr
& Process && Partner Process && Amp.\ Error && SU(3) Breaking & \cr
height2pt&\omit&&\omit&&\omit&&\omit&\cr
\noalign{\hrule}
height2pt&\omit&&\omit&&\omit&&\omit&\cr
& $\bd \to D^+ D^-$ && $\bs \to D_s^+ D_s^-$ && 0 && 
${\displaystyle f_{D_s} \over \displaystyle f_D} \, {\displaystyle
F_0^{\bs \to D_s} \over \displaystyle F_0^{\bd \to D}}$ & \cr
& \omit && $\bd \to D_s^+ D^-$ && $\sim 5\%$ &&
$f_{D_s}/f_D$ & \cr
& \omit && $B_u^+ \to D_s^+ {\bar D}^0$ && $\sim 5\%$ && $f_{D_s}/f_D$
& \cr
height2pt&\omit&&\omit&&\omit&&\omit&\cr
\noalign{\hrule}
height2pt&\omit&&\omit&&\omit&&\omit&\cr
& $\bd \to \pi^+ \pi^-$ && $\bs \to K^0 {\bar K}^0$ && $\sim 5\%$ &&
${\displaystyle f_K \over \displaystyle f_\pi} \, {\displaystyle
F_0^{\bs \to {\bar K}} \over \displaystyle F_0^{\bd \to \pi}}$ & \cr
& \omit && $B_u^+ \to K^0 \pi^+$ &&$\sim 10\%$ && $f_K/f_\pi$ & \cr
height2pt&\omit&&\omit&&\omit&&\omit&\cr
\noalign{\hrule}
height2pt&\omit&&\omit&&\omit&&\omit&\cr
& $\bd \to K^0 {\bar K}^0$ && $\bs \to K^0 {\bar K}^0$ && 0 &&
${\displaystyle F_0^{\bs \to {\bar K}} \over \displaystyle F_0^{\bd \to K}}$ & \cr
& \omit && $B_u^+ \to K^0 \pi^+$ && $\sim 5\%$ && 
${\displaystyle F_0^{\bd \to \pi} \over \displaystyle F_0^{\bd \to K}}$ & \cr
height2pt&\omit&&\omit&&\omit&&\omit&\cr
\noalign{\hrule}
height2pt&\omit&&\omit&&\omit&&\omit&\cr
& $\bd\to\rho^0\rho^0$ && $\bd\to K^{*0}\rho^0$ && $\sim 10\%$ &&
$g_\rho/g_{K^*}$ & \cr
height2pt&\omit&&\omit&&\omit&&\omit&\cr
\noalign{\hrule}
height2pt&\omit&&\omit&&\omit&&\omit&\cr
& $\bs \to {\bar K}^{*0} \rho^0$ && $\bd \to K^{*0} \rho^0$ && $\sim
5\%$ && 
${\displaystyle g_\rho \over \displaystyle g_{K^*}} \, {\displaystyle
F^{\bs \to {\bar K}^*} \over \displaystyle F^{\bd \to \rho}}$ & \cr
height2pt&\omit&&\omit&&\omit&&\omit&\cr
\noalign{\hrule}
height2pt&\omit&&\omit&&\omit&&\omit&\cr
& $\bs \to \phi {\bar K}^{*0}$ && $\bd\to\phi K^{*0}$ && 0 &&
${\displaystyle g_\phi \over \displaystyle g_{K^*}} \, {\displaystyle
F^{\bd \to K^*} \over \displaystyle F^{\bs \to \phi}}$ & \cr
& \omit && $B_u^+ \to \phi K^{*+}$ && $\sim 1\%$ &&
${\displaystyle g_\phi \over \displaystyle g_{K^*}} \, {\displaystyle
F^{\bd \to K^*} \over \displaystyle F^{\bs \to \phi}}$ & \cr
& \omit && $\bs\to\phi\phi$ && $\sim 5\%$ && 
$g_\phi/g_{K^*}$ & \cr
height2pt&\omit&&\omit&&\omit&&\omit&\cr
\noalign{\hrule}}}
\caption{$B$ decays and their partner processes which can be used to
obtain CP phase information with Method I or II. The theoretical
uncertainties due to the neglect of certain amplitudes (Amp.\ Error)
and SU(3) breaking are shown. For most decays, the expressions for the
SU(3)-breaking error also include (known) mass factors. For the
vector-vector final states, the ``form factor'' $F^{B\to f}$ is
symbolic only: see Eqs.~(\protect\ref{su3Kstarrhoapprox}) and
(\protect\ref{su3Kstarphiabc}) for the exact expressions.}
\label{summarytable}
}

For each pair, there are two main sources of theoretical uncertainty.
First, there is the ``amplitude error,'' which corresponds to the
error due to the neglect of certain diagrams, usually of
exchange/annihilation type. These diagrams have been estimated to be
at most $O({\bar\lambda}^2) \sim 5\%$ of the leading decay amplitude
[Eqs.~(\ref{buudhierarchy}) and (\ref{buushierarchy})]. In most cases
one diagram must be neglected, leading to an amplitude error of $\sim
5\%$. However, there are some decay pairs for which this error is zero
(no diagrams neglected) or $\sim 10\%$ (two diagrams neglected).

As noted earlier, in some approaches to hadronic $B$ decays
\cite{BBNS,PQCD}, diagrams corresponding to exchange/annihilation may
be enhanced for final states involving pseudoscalars due to ``chiral
enhancement'' factors such as those found in
Eq.~(\ref{chiralenhance}). If true --- and this can be tested
experimentally --- then the amplitude errors given in
Table~\ref{summarytable} may be underestimated for certain decays. In
this case, one can avoid chiral enhancements by considering only
decays with vector-vector final states. For such decays, an angular
analysis will be necessary to separate out the three different
helicity states. Method I can then be applied to a single such state,
while two helicity states are required for Method II.

The second source of theoretical uncertainty is the breaking of flavor
SU(3) symmetry in the ratio of Eq.~(\ref{assumption}), which is
expected to be $O(m_s/\Lambda_{QCD}) \sim 25\%$. In order for our
methods to be useful, this SU(3) uncertainty must be reduced. We have
estimated the SU(3) breaking for each pair of decays, and it appears
in the last column of Table~\ref{summarytable}. For five decay pairs,
the SU(3) breaking is given solely by a ratio of decay constants,
sometimes multiplied by a function of (known) masses. Many of these
decay constants have been measured experimentally. For pseudoscalars,
we have $f_\pi = 131$ MeV and $f_K = 160$ MeV \cite{pdg}; for vectors,
$g_{\rho}=209$ MeV, $g_{K^*}=218$ MeV and $g_{\phi}=221$ MeV
\cite{benekeneubert}.  The ratios $f_K/f_\pi$, $g_\rho/g_{K^*}$ and
$g_\phi/g_{K^*}$ are therefore all known, with small errors. There are
also experimental values for another ratio, $f_{D_s}/f_D$, but the
errors are huge \cite{pdg}.  Instead, one can take the value of this
ratio from the lattice: $f_{D_s}/f_D = 1.22 \pm 0.04$ \cite{lattice},
which has a very small error. Thus, for the five decay pairs in which
the leading SU(3)-breaking term is given by a ratio of decay
constants, the unknown theoretical SU(3) uncertainty in
Eq.~(\ref{assumption}) is only a second-order effect. In this case,
one can use Method I to extract CKM phase information from these pairs
of decays. {}From an experimental point of view, these are the favored
pairs of decay modes. In Sec.~6, using the latest data, we show
explictly how Method I can be applied to $\bd \to \pi^+ \pi^-$ and
$B_u^+ \to K^0 \pi^+$ decays to extract $\gamma$.

For the remaining seven decay pairs, the SU(3) breaking is given by an
expression involving form factors. Unfortunately, these form factors
cannot be calculated yet in QCD, and so one must find a way of
estimating them in order to reduce the size of SU(3) breaking. One way
is to simply rely on model calculations. However, it is difficult to
argue that the theoretical error is smaller than the canonical size of
SU(3) breaking, $\sim 25\%$.

In some cases, the ratio of form factors can be {\it measured} using
the partner processes. For example, the ratio of amplitudes for $\bd
\to D_s^+ D^-$ and $\bs \to D_s^+ D_s^-$ is proportional to $F_0^{\bd
\to D}/F_0^{\bs \to D_s}$. Thus, to a good approximation, the
measurement of the rates for these processes allows us to measure the
desired ratio of form factors. Similarly, $F_0^{\bs \to
\bar{K}}/F_0^{\bd \to \pi}$ and $F^{\bd \to \bar{K}^*}/F^{\bs \to
\phi}$ can be obtained from measurements of the partner processes $\bs
\to K^0 {\bar K}^0$ and $B_u^+ \to K^0 \pi^+$, and $\bd \to \phi
K^{*0}$ and $\bs \to \phi\phi$, respectively. One can therefore
extract ratios of form factors from the measurement of the branching
ratios for two $\btos$ penguin decays, up to the theoretical error
incurred by neglecting exchange- and annihilation-type diagrams.

Another way to measure the ratio of form factors is to consider the
heavy-quark limit $m_{b,c} \to \infty$. In this case, the ratio of
form factors in $B$ decays is related to a similar ratio in $D$
decays. That is,
\beq
{F^{\bs \to f_1} \over F_{\bd \to f_2}} = {F^{\bar{D}_s \to f_1} 
\over F_{\bar{D}
\to f_2}} ~,~~
{F^{\bd \to f_1} \over F_{\bd \to f_2}} = {F^{\bar{D} \to f_1} 
\over F_{\bar{D} \to
f_2}} ~.
\eeq
The $D$ form-factor ratios can be measured in semileptonic $D$
decays. The correction to the relation between $B$ and $D$ form
factors is $O(1/m_{b,c})$. Most of the form-factor ratios in
Table~\ref{summarytable} can in principle be measured in this way.

If it is not possible to get information about the form-factor ratios
using the above methods, one can reduce the error due to SU(3)
breaking by using Method II. In this case one uses two decay pairs
related by SU(3). The double ratio of decay amplitudes then provides
the necessary input to extract CP-phase information.

Several decay pairs in Table~\ref{summarytable} involve pseudoscalar
final states. In applying Method II to these decays, it is tempting to
take as the second decay pair the corresponding vector-vector decay.
However, for $B \to light$ transition, this will not work well: there
are no relations between $B \to P$ and and $B \to V$ form factors.
Thus, though intuitively we expect some cancellation when we consider
the double ratio of form factors, the amount of cancellation is
model-dependent.

There are two exceptions to this. First, in $B \to heavy$ transitions
all the $B \to P$ and $B \to V$ form factors can be expressed in terms
of a single Isgur-Wise function in the $m_{b,c} \to \infty$ limit.
Thus, for the decays $\bd \to D^{(*+)} D^{(*-)}$ and $\bs \to
D_s^{(*+)} D_s^{(*-)}$, we can use Method II with pseudoscalar and
vector final states, and the SU(3) breaking will cancel in the
$m_{b,c} \to \infty$ limit. The SU(3)-breaking effects from finite
$m_{b,c}$ will be suppressed by $m_s/m_{b,c}$.

The second exception is the decay pair $\bd \to K^0 {\bar K}^0$ and
$\bs \to K^0 {\bar K}^0$. This case is special because the final state
is the same for both processes. Here the ratio of pseudoscalar form
factors differs from unity (for a symmetric kaon LCD) only due to the
initial-state correction which arises because of the difference in the
light degrees of freedom in the $\bd$ and $\bs$ mesons. A similar
correction will appear in the ratio of vector form factors. Indeed,
using the QCD sum rule model of form factors, the initial-state
correction is the same for both $P$ and $V$ form factors, so that it
cancels in the double ratio. We therefore find that, to leading order,
there is no SU(3) breaking in Method II \cite{BKKbar}.

Apart from these two cases, Method II works best when two polarization
states of the $VV$ final state are used. For the remaining ($B \to
light$) decays, to the extent that Eq.~(\ref{leet}) is valid, the
three polarization amplitudes in the principal $B\to VV$ decay, as
well as its partner process, are expressible in terms of a single
universal form factor.  Thus, all dependence on form factors vanishes
in the double ratio, leaving only a theoretical uncertainty at the
level of second-order SU(3) breaking. (Note: we have found that a
$\pm$30\% difference in $\zeta_{\|}$ and $\zeta_{\perp}$ in
Eq.~(\ref{leet}) does not significantly affect the calculation of
SU(3) breaking.)

However, several caveats must be pointed out. First, it is not certain
that the assumptions behind the relations in Eq.~(\ref{leet}) are
valid. It is necessary to measure the form factors in semileptonic $B$
decays to test the validity of Eq.~(\ref{leet}). If these relations
are not found to be valid, then, while we expect that much of the
SU(3) breaking in the double ratio will cancel, this is not guaranteed
by any symmetry principle. (However, such cancellations occur to
varying degrees in most form-factor models.) 

It is also possible to use Method II for decay pairs in which the main
SU(3)-breaking uncertainty is due to decay constants. However, this is
unnecessary, since the values of the decay constants are known.

It is clear that Method II will only apply to decays which have more
than one polarization state. Thus, it will not apply to the decay $B
\to \rho^+ \rho^-$, which has been found to be dominated by the
longitudinal polarization \cite{exptpol}. On the other hand, the
measurement of the polarization amplitudes in $\bd \to \phi K^*$
indicates a sizeable transverse component \cite{exptpol}, so that
Method II should be feasible in this case.

Finally, we address the question of which of the twelve $B$ decay
pairs is most promising for extracting CKM phase information. As noted
earlier, for five pairs the leading theoretical uncertainty is given
by a ratio of decay constants only, and is therefore already known. Of
these, one pair --- $\bs \to \phi {\bar K}^{*0}$ and $\bs\to\phi\phi$
--- involves $\bs$ decays. The measurements of these decays will
eventually be made by future hadron colliders, but this will take a
number of years. A second pair --- $\bd\to\rho^0\rho^0$ and $\bd\to
K^{*0}\rho^0$ --- requires an angular analysis. In addition, the
branching ratios for these decays are expected to be small, so these
measurements will also take some time. The most promising applications
of our methods therefore involve the remaining three decays. For one
of these --- $\bd \to \pi^+ \pi^-$ and $B_u^+ \to K^0 \pi^+$ --- data
is already available. We present the analysis of this pair of decays
in Sec.~6. Measurements of CP violation in the other decay pairs ---
$\bd \to D^+ D^-$ and $\bd \to D_s^+ D^-$ \cite{BDDbar} or $B_u^+ \to
D_s^+ {\bar D}^0$ --- will likely be made soon.

To conclude, we have found a number of pairs of $B$ decays, related by
flavor SU(3), whose measurements can be combined to furnish
information about CP phases. The main source of theoretical
uncertainty in relating the decays $B^0 \to M_1 M_2$ and $B' \to M'_1
M'_2$ is SU(3) breaking, which is typically $O(m_s/\Lambda_{QCD}) \sim
25\%$. We have given several ways of removing the leading-order SU(3)
uncertainty. This leaves a second-order error of $\sim 5\%$. Depending
on the decay pair chosen, and how one combines the various theoretical
uncertainties, the theoretical error is between 5\% and 15\%.

\section{Extracting $\gamma$ from $\bd \to \pi^+ \pi^-$ and 
$B_u^+ \to K^0 \pi^+$}

In this section, using the latest data, we show explictly how our
Method I can be applied to $\bd \to \pi^+ \pi^-$ and $B_u^+ \to K^0
\pi^+$ decays to obtain the CP phase $\gamma$. (This is basically an
update of Ref.~\cite{GroRosner}.) As we will see, although the method
is in principle feasible, the present experimental errors are too
large to allow a determination of $\gamma$.

We define the CP asymmetries as follows:
\beq
C_{\pi\pi} \equiv { \left( |A|^2 - |{\overline A}|^2 \right) \over
  \left( |A|^2 + |{\overline A}|^2 \right) } = {a_{dir} \over B} ~,~~
S_{\pi\pi} \equiv { 2 {\rm Im} \left( e^{-2i \beta} A^* {\overline
A} \right) \over \left( |A|^2 + |{\overline A}|^2 \right) } = {\aI
\over B} ~.
\eeq
In terms of these quantities, Eq.~(\ref{aRdef}) can be written
\beq
S_{\pi\pi}^2 + C_{\pi\pi}^2 + D_{\pi\pi}^2 = 1 ~,
\label{SCDrel}
\eeq
where $D_{\pi\pi} \equiv \aR/B$ \cite{GroRosner2}. Then
Eq.~(\ref{gammacond}) becomes
\beq
{\Act^2 \over B} = { D_{\pi\pi} \cos(2\beta + 2\gamma) - S_{\pi\pi}
\sin(2\beta + 2\gamma) - 1 \over \cos 2\gamma - 1} ~.
\label{gammacond2}
\eeq
According to Method I, $\lambda \Actp$ is related to $\Act$. As shown
earlier, the leading-order SU(3)-breaking effect is expressible in
terms of decay constants only:
\beq
{\lambda \Actp \over \Act} = {f_K \over f_\pi} = 1.22 ~.
\label{assumption'}
\eeq

We now expand $\cos(2\beta + 2\gamma)$ and $\sin(2\beta + 2\gamma)$ in
Eq.~(\ref{gammacond2}), and write $\cos 2\gamma = \sqrt{1 - \sin^2 2
\gamma}$. This gives a quadratic equation for $\sin 2 \gamma$:
\beq
{\tilde A} \sin^2 2\gamma + {\tilde B} \sin 2\gamma + {\tilde C} = 0 ~,
\label{quadratic}
\eeq
where
\bea
{\tilde A} & = & \left( D_{\pi\pi} \sin 2\beta + S_{\pi\pi} \cos
2\beta \right)^2 + \left( D_{\pi\pi} \cos 2\beta - S_{\pi\pi} \sin
2\beta - {\Act^2 \over B} \right)^2 ~, \nn\\
{\tilde B} & = & 2 \left( D_{\pi\pi} \sin 2\beta + S_{\pi\pi} \cos
2\beta \right) \left( 1 - {\Act^2 \over B} \right) ~, \nn\\
{\tilde C} & = & \left( 1 - {\Act^2 \over B} \right)^2 - \left(
D_{\pi\pi} \cos 2\beta - S_{\pi\pi} \sin 2\beta - {\Act^2 \over B}
\right)^2 ~.
\label{tildeparams}
\eea

In order to extract $\gamma$, we need the results for the latest
measurements of the above quantities. According to Ref.~\cite{pdg}, we
have
\beq
\sin 2\beta = 0.73 \pm 0.05 ~,~~ \cos 2\beta = 0.68 \pm 0.04 ~,
\eeq
where we have assumed the value of $2\beta$ consistent with the
SM. (If $\cos 2\beta = -0.68$ is taken, then this would already be
evidence for physics beyond the SM.) Ref.~\cite{pdg} also gives
\beq
BR(\bd \to \pi^+ \pi^-) = (4.8 \pm 0.5) \times 10^{-6} ~.
\eeq
Writing all amplitudes in units of branching ratios of $10^{-6}$, with
the $\bd$ lifetime, this gives
\beq
B = \sqrt{4.8 \pm 0.5} = 2.2 \pm 0.1 ~.
\eeq

For $B^+ \to \pi^+ K^0$, the most recent papers by Belle and BaBar
give \cite{BelleBaBar}
\bea
{\rm Belle:} ~~ BR(B^+ \to \pi^+ K^0) & = & (2.20 \pm 0.19~(stat) \pm
0.11~(syst)) \times 10^{-5} ~, \nn\\
{\rm BaBar:} ~~ BR(B^+ \to \pi^+ K^0) & = & (2.23 \pm 0.17~(stat) \pm
0.11~(syst)) \times 10^{-5} ~.
\eea
Using these numbers, we find
\beq
BR(B^+ \to \pi^+ K^0) = (2.22 \pm 0.15) \times 10^{-5} ~,
\eeq
so that
\beq
{\Actp}^2 = (22.2 \pm 1.5)/r_\tau ~,
\eeq
where $r_\tau = \tau_{B^+}/\tau_{\bd} = 1.085$ \cite{pdg}. This gives
\beq
{\Act^2 \over B} = {\lambda^2 \over B} \left( {f_\pi \over f_K}
\right)^2 {\Actp}^2 = 0.31 \pm 0.03 ~.
\label{Arels}
\eeq
It is here that the theoretical error should be added. This error
comes from the neglect of small diagrams, as well as SU(3) breaking.
For the moment, we do not include any theoretical error; we comment on
this below.

For the CP asymmetries, we average the latest Belle and BaBar data
\cite{BelleBaBarCP}:
\bea
S_{\pi\pi} & = & -0.70 \pm 0.19 ~, \nn\\
C_{\pi\pi} & = & -0.42 \pm 0.13 ~,
\eea
giving a value
\beq
D_{\pi\pi} = \pm ( 0.58 \pm 0.24 ) ~.
\eeq

Taking the central values of the above numbers, we find:
\bea
D_{\pi\pi} > 0 &:& ~ \sin 2\gamma = 0.10 \pm 0.57 i ~, \nn\\
D_{\pi\pi} < 0 &:& ~ \sin 2\gamma = 0.87 ~, 0.59 ~.
\label{gammavalues}
\eea
A real value for $\sin 2\gamma$ is found only for $D_{\pi\pi} < 0$.

However, we must include the errors on the various measurements.
Considering only experimental errors, we find (for $D_{\pi\pi} < 0$)
\beq
\sin 2\gamma = 0.87 \pm 1.45 ~,~~ 0.59 \pm 1.42 ~.
\eeq
Recall that a value for $\sin 2\gamma$ gives four values for $\gamma$.
Thus, it is clear that, within this method, present measurements place
very few constraints on the value of $\gamma$.

We now turn to the question of theoretical error. In relating the
penguin amplitudes for $\bd \to \pi^+ \pi^-$ and $B_u^+ \to K^0
\pi^+$, we have neglected two small diagrams: $PA$ and $\pewc$ (or
$\pewcp$). Each of these is expected to be $\sim 5\%$ of the dominant
$P$ (or $P'$) amplitude. (We have also neglected the annihilation
diagram $A'$, but this is expected to be tiny compared to $P'$
[Eq.~(\ref{buushierarchy})].) The leading-order SU(3)-breaking term is
known with little error ($f_\pi/f_K$), but the second-order effect is
unknown. The size of this effect is estimated to be $\sim 5\%$. Thus,
the total theoretical error in relating $\Act$ to $\Actp$ is between
8\% (errors added in quadrature) and 15\% (errors added
linearly). Since the method involves the squares of the amplitudes,
the theoretical error in Eq.~(\ref{Arels}) is between 15\% and 30\%.

What is the effect on the extraction of $\sin 2\gamma$? Setting all
experimental errors to zero, we consider a theoretical error of 20\%. 
In this case we find
\beq
\sin 2\gamma = 0.87 \pm 0.50 ~,~~ 0.59 \pm 0.50 ~.
\eeq
The effect on $\sin 2\gamma$ is obviously very large. The reason is
that $\sin 2\gamma$ depends on ${\tilde B}^2 - 4 {\tilde A}{\tilde C}$
[see Eq.~(\ref{quadratic})]. For the particular values of the
measurements of $\bd \to \pi^+ \pi^-$ and $B_u^+ \to K^0 \pi^+$, it
turns out that there is a large cancellation between ${\tilde B}^2$
and $4 {\tilde A}{\tilde C}$, so that the percentage error here is
enormous. This leads to the large error on $\sin 2\gamma$.

The lesson here is that, although the theoretical error in relating
$\Act$ and $\Actp$ is indeed between 5\% and 15\% for all decay pairs,
its effect on the extraction of CP phases cannot be predicted. It
depends sensitively on the values found for the various measurements
of the $B$ decay pairs.

\newpage
\section{Conclusions}

We have examined in detail a method for obtaining CP phase information
from pairs of $B$ decays. The method works as follows. Consider a
decay $B^0 \to M_1 M_2$ which involves a $\btod$ penguin amplitude,
and in which both $B^0$ and $\Bbar$ mesons can decay to $M_1 M_2$.
Now consider a second decay $B' \to M'_1 M'_2$, related to the first
by flavor SU(3), which receives a significant $\btos$ penguin
contribution and which is dominated by a single amplitude. Using the
fact that the two decays are related, one has enough information to
extract CP phases from the measurements of the two decays. Depending
on how one parametrizes the decay amplitudes, either $\alpha$ or
$\gamma$ can be obtained (assuming that $\beta$ is independently
measured).

There are two sources of theoretical error in this method. First, it
may be necessary to neglect certain diagrams in order to find a decay
pair related by SU(3). We neglect only amplitudes which are expected
to be small (e.g.\ exchange, annihilation, color-suppressed
electroweak penguins); only pairs for which this error is at most 10\%
are considered. Second, SU(3) breaking, which is typically
$O(m_s/\Lambda_{QCD}) \sim 25\%$, must be taken into account. We show
that, in general, one can eliminate the leading-order SU(3)-breaking
effects, leaving only a second-order error $\sim 5\%$. This occurs in
one of two ways. First, for certain decay pairs the leading-order
effect depends only on (known) decay constants. Second, for those
decays for which this effect depends on (unknown) form factors, there
are several ways to proceed. One can measure the ratio of form
factors, either in $B$ or in $D$ decays. Alternatively, by taking a
double ratio of related SU(3) decays, the leading-order SU(3)-breaking
effect cancels.

Our results are shown in Table~\ref{summarytable}. We find twelve
decay pairs to which this method can be applied. Some of these ---
$\bd \to D^+ D^-$ and $\bd \to D_s^+ D^-$, $\bd \to D^+ D^-$ and $\bs
\to D_s^+ D_s^-$, $\bd \to \pi^+ \pi^-$ and $B_u^+ \to K^0 \pi^+$, and
$\bd \to K^0 {\bar K}^0$ and $\bs \to K^0 {\bar K}^0$ --- have been
examined in the literature. However, the other pairs are new.
Depending on which decay pair is used, and how one combines the
various theoretical uncertainties, the total theoretical error in
relating the amplitudes for the decays $B^0 \to M_1 M_2$ and $B' \to
M'_1 M'_2$ is between 5\% and 15\%. However, the error in the
extraction of CP phases cannot be predicted -- it depends sensitively
on the values of the measurements of the two $B$ decays. The most
promising decay pairs are $\bd \to \pi^+ \pi^-$ and $B_u^+ \to K^0
\pi^+$, for which data is already available, and $\bd \to D^+ D^-$ and
$\bd \to D_s^+ D^-$ or $B_u^+ \to D_s^+ {\bar D}^0$.  CP-violating
measurements of these latter decays will probably be made very soon.

\bigskip
\noindent
{\bf Acknowledgements}:
We thank J.L. Rosner for helpful conversations, and M. Gronau and
J.L. Rosner for bringing Ref.~\cite{GroRosner} to our attention. This
work was financially supported by NSERC of Canada.



\begin{thebibliography}{99}

\bibitem{pdg} K. Hagiwara {\it et al.}  [Particle Data Group
Collaboration], \prd{66}{2002}{010001},
http://pdg.lbl.gov/pdg.html.

\bibitem{CPreview} For a review, see, for example, {\it The BaBar
  Physics Book}, eds.\ P.F. Harrison and H.R. Quinn, SLAC Report 504,
  October 1998.

\bibitem{su3} There are too many SU(3) references to present a
detailed list here. First analyses can be found in M. Gronau,
O.F. Hern\'andez, D. London and J.L. Rosner, \prd{50}{1994}{4529},
\prd{52}{1995}{6374}, \prd{52}{1995}{6356}; M. Gronau, J.L. Rosner and
D. London, \prl{73}{1994}{21}; O.F. Hern\'andez, D. London, M. Gronau
and J.L. Rosner, \plb{333}{1994}{500}.

\bibitem{BKKbar} A. Datta and D. London, \plb{533}{2002}{65}. The
  decay $B^0_{d,s} \to K^{*0} {\bar K}^{*0}$ has also been discussed
  in R. Fleischer, \prd{60}{1999}{073008}.

\bibitem{BDDbar} A. Datta and D. London, \plb{584}{2004}{81}.

\bibitem{LSS} The fact that one needs theoretical input to obtain CP
phase information from $\btod$ penguin decays was noted in D. London,
N. Sinha and R. Sinha, \prd{60}{1999}{074020}.

\bibitem{BBNS} M. Beneke, G. Buchalla, M. Neubert and C.T. Sachrajda,
\prl{83}{1999}{1914}, \npb{591}{2000}{313}, \npb{606}{2001}{245}.

\bibitem{GroRosner} M. Gronau and J.L. Rosner, \prd{66}{2002}{053003},
[Erratum-ibid.\ D {\bf 66}, 119901 (2002)]

\bibitem{eta} See, for example, A. Datta, X.-G. He and S. Pakvasa,
\plb{419}{1998}{369}; A. Datta, H.J. Lipkin and P.J. O'Donnell,
\plb{529}{2002}{93}.

\bibitem{Fleischer1} R. Fleischer, \plb{459}{1999}{306}.

\bibitem{BsKKSU3break} M. Beneke, eConf {\bf C0304052}, FO001 (2003)
[arXiv:hep-ph/0308040].

\bibitem{Fleischer2} R.~Fleischer, \epjc{10}{1999}{299}.

\bibitem{firstSU3} M. Gronau, O.F. Hern\' andez, D. London and
J.L. Rosner, \prd{50}{1994}{4529}.

\bibitem{EWPs} M. Gronau, O.F. Hern\' andez, D. London and
J.L. Rosner, \prd{52}{1995}{6374}.

\bibitem{GSW} S. Bertolini, F. Borzumati, and A. Masiero,
\prl{59}{1987}{180}; N.G. Deshpande, P. Lo, J. Trampetic, G. Eilam,
and P. Singer, \prl{59}{1987}{183}; B. Grinstein, R. Springer, and
M. Wise, \plb{202}{1988}{138}; \npb{339}{1990}{269}.

\bibitem{PQCD} Y.Y.~Keum, H.-n.~Li and A.I.~Sanda, \plb{504}{2001}{6}.

\bibitem{luorosner} J.L. Rosner, \prd{42}{1990}{3732}; Z.~Luo and
J.L.~Rosner, \prd{64}{2001}{094001}. See also A.~Abd El-Hady, A.~Datta
and J.P.~Vary, \prd{58}{1998}{014007}; A.~Abd El-Hady, A.~Datta,
K.S.~Gupta and J.P.~Vary, \prd{55}{1997}{6780}.

\bibitem{BuraseffH} See, for example, G. Buchalla, A.J. Buras and
  M.E. Lautenbacher, {\it Rev.\ Mod.\ Phys.} {\bf 68}, 1125 (1996),
  A.J. Buras, ``Weak Hamiltonian, CP Violation and Rare Decays,'' in
  {\it Probing the Standard Model of Particle Interactions}, ed.\
  F. David and R. Gupta (Elsevier Science B.V., 1998), pp.\ 281-539.

\bibitem{BSW} M. Bauer, B. Stech and M, Wirbel, \zpc{34}{1987}{103}.

\bibitem{su3isgur} E. Jenkins and M.J. Savage, \plb{281}{1992}{31};
A.F. Falk, \plb{305}{1993}{268}.

\bibitem{lattice} D. Becirevic, invited talk at {\it 2nd Workshop on
the CKM Unitarity Triangle}, Durham, England, April 2003,
hep-ph/0310072.

\bibitem{lukewise} Z.~Ligeti, M.E.~Luke and M.B.~Wise,
\plb{507}{2001}{142}.

\bibitem{charming} M. Ciuchini, E. Franco, G. Martinelli, M. Pierini
and L. Silvestrini, \plb{515}{2001}{33}.

\bibitem{Ballv} P. Ball and V.M. Braun, \newprd{58}{1998}{094016};
P. Ball, \jhep{09}{1998}{005}; P. Ball and V.M. Braun,
\npb{543}{1999}{201}.

\bibitem{pionLCD} E.M. Aitala {\it et al.}  [E791 Collaboration],
\prl{86}{2001}{4768}.

\bibitem{dattalondonTP} A. Datta and D. London, arXiv:hep-ph/0303159.

\bibitem{Charles} J. Charles et al., \prd{60}{1999}{014001}.

\bibitem{benekeneubert} M. Beneke and M. Neubert,
\npb{675}{2003}{333}.

\bibitem{exptpol} J. Fry, talk given at Lepton-Photon 2003, Fermilab,
Batavia, Illinois, 11-16 August 2003,
http://conferences.fnal.gov/lp2003.

\bibitem{GroRosner2} M. Gronau and J.L. Rosner,
\prd{65}{2002}{093012}.

\bibitem{BelleBaBar} Y.~Chao {\it et al.}  [Belle Collaboration],
arXiv:hep-ex/0311061; B.~Aubert {\it et al.}  [BABAR Collaboration],
arXiv:hep-ex/0312055.

\bibitem{BelleBaBarCP} K.~Abe {\it et al.}  [Belle Collaboration],
arXiv:hep-ex/0401029; H. Jawahery, talk given at Lepton-Photon 2003,
Fermilab, Batavia, Illinois, 11-16 August 2003,
http://conferences.fnal.gov/lp2003. Averages taken from A.J. Buras,
R.~Fleischer, S.~Recksiegel and F.~Schwab, arXiv:hep-ph/0402112.

\end{thebibliography}
\end{document}